\pdfoutput=1

\documentclass[11pt]{article}
\usepackage{jheppub}

\usepackage{amsmath}
\usepackage{amssymb}
\usepackage{graphicx,color,slashed,braket}
\usepackage{bbm}
\usepackage{ifpdf}
\usepackage{comment}
\usepackage{soul}

\usepackage{booktabs}
\usepackage{multirow}
\usepackage{makecell}
\usepackage{float}
\usepackage{verbatim}
\usepackage{caption}
\usepackage{cancel}
\usepackage{enumerate}
\usepackage{tcolorbox}
\usepackage[font=small,labelfont=bf]{caption}

\usepackage{color}
\definecolor{dark-gray}{gray}{0.20}
\definecolor{gray}{gray}{0.30}
\definecolor{light-gray}{gray}{0.80}
\definecolor{dark-red}{rgb}{0.7,0,0}
\definecolor{dark-green}{rgb}{0.1,0.4,0}
\definecolor{dark-blue}{rgb}{0.3,0.3,0.7}
\definecolor{light-blue}{rgb}{0.8,0.8,1}

\usepackage{tikz}
\usetikzlibrary{calc}

\usepackage{hyperref}
\usepackage{setspace}
\hypersetup{
	colorlinks=true,
	linkcolor=dark-blue,
	citecolor=dark-red,
	urlcolor=dark-green,
	linktoc=page
}
\newcommand{\be}{\begin{equation}}
\newcommand{\ee}{\end{equation}}

\def\be{\begin{equation}}
\def\ee{\end{equation}}
\def\bea{\begin{eqnarray}}
\def\eea{\end{eqnarray}}

\newcommand{\e}{\mathrm{e}}

\newcommand{\dd}{\mathrm{d}}

\newcommand{\vol}{\text{vol}}

\renewcommand{\Im}{\text{Im}\,}
\renewcommand{\Re}{\text{Re}\,}

\allowdisplaybreaks

\title{Broken and restored: a holographic constraint for AdS vacua with orbifolds}

\preprint{UUITP-06/26}
\author[1]{Filippo Revello}
\author[2]{\& Vincent Van Hemelryck}

\affiliation[1]{Instituut voor Theoretische Fysica and Leuven Gravity Institute, KU Leuven, Celestijnenlaan 200D, B-3001 Leuven, Belgium}
\affiliation[2]{Department of Physics and Astronomy, Uppsala University, Box 516, SE-75120, Uppsala, Sweden}

\emailAdd{filippo.revello@kuleuven.be}
\emailAdd{vincent.vanhemelryck@physics.uu.se}

\notoc

\abstract{It has been suggested that families of weakly-coupled AdS vacua with a large-$N$ holographic dual must satisfy non-trivial consistency requirements, which amount to the vanishing of certain cubic couplings, corresponding to (super-)extremal arrangements of scalar operators. While this constraint is known to hold in the simplest incarnation of the DGKT scenario in massive type IIA string theory, i.e. on the $\mathbb{Z}_3\times \mathbb{Z}_3$ orbifold, we find that it is generically violated for type II AdS$_3$ and AdS$_4$ vacua arising from $\mathbb{Z}_2 \times \mathbb{Z}_2 \times \mathbb{Z}_2$ and $\mathbb{Z}_2 \times \mathbb{Z}_2$ orbifolds respectively, including scale-separated solutions and DGKT--CFI-type models. In most cases, however, this can be cured by enlarging the orbifold group to a suitable (non-abelian) extension that projects out precisely those scalar operators that would otherwise participate in the constrained cubic couplings. 
Our results suggest that consistency of the putative holographic dual imposes a non-trivial restriction on the compactification geometry, indicating in particular that O-planes cannot wrap cycles in distinct homology classes.}
\begin{document}

\maketitle

\newpage
\tableofcontents
\vspace{0.5cm}
\hrule
\section{Introduction}

A central challenge in string theory, and a key focus of the swampland programme, is to test the ultraviolet (UV) consistency of gravitational effective field theories (EFTs) within quantum gravity. In recent years, numerous criteria have been proposed and explored, with significant success for EFTs with vanishing cosmological constant, largely due to insights from Minkowski compactifications of string theory.
In contrast, for EFTs with a nonzero cosmological constant, which require moduli stabilisation, many of the existing swampland conjectures do not directly apply. More precisely, EFTs with a cosmological constant must exhibit scale separation, i.e. the cosmological constant must be parametrically smaller than the UV cut-off set by the Kaluza--Klein scale. Although examples of such compactifications with negative cosmological constant exist in string theory, their validity remains the subject of considerable debate.
These constructions achieve moduli stabilisation through intricate configurations of fluxes and (intersecting) orientifold planes, typically relying on simplifying approximations. The most extensively studied example is the four-dimensional DGKT--CFI construction in massive type IIA \cite{DeWolfe:2005uu,Camara:2005dc}. Related AdS$_4$ vacua in massless type IIA \cite{Cribiori:2021djm,Carrasco:2023hta,VanHemelryck:2024bas} have also been investigated, along with AdS$_3$ vacua in both type IIA \cite{Farakos:2020phe,Farakos:2023nms,Farakos:2025bwf,Tringas:2026ncg} and type IIB \cite{VanHemelryck:2025qok,Miao:2025rgf}, as well as their heterotic counterparts \cite{Tringas:2025bwe}. All of these solutions preserve minimal or no supersymmetry, and refs.~\cite{Cribiori:2020use,Cribiori:2022trc,Montero:2022ghl,Cribiori:2024jwq} suggest that scale separation cannot be achieved in setups with extended supersymmetry, possibly due to the existence of a non-trivial R-symmetry in the dual CFT.\footnote{Some 3d vacua with putative 2d $\mathcal{N}=(1,1)$ CFT duals were recently discussed in \cite{Cribiori:2026caf}.}
For a long time, the non-trivial backreaction of (intersecting) orientifold planes was the primary source of concern. However, a series of works \cite{Baines:2020dmu,Junghans:2020acz,Marchesano:2020qvg, Emelin:2022cac,VanHemelryck:2024bas} has shown that this backreaction remains small at first order in perturbation theory, thereby justifying the smeared orientifold approximation.\footnote{It has been suggested that the DGKT vacua may be incompatible with the membrane version of the Weak Gravity Conjecture, whereas this issue does not arise for the vacua of \cite{Montero:2024qtz}.}
Together with higher-derivative corrections, this prompts the question of whether it would even be possible to \emph{definitively} establish the validity of such solutions in a direct manner, which relies on a perturbative EFT logic. Even if the required approximations were found to be valid at arbitrarily high order in perturbation theory, it would always be possible for an obstruction to appear at the next order, or even non-perturbatively. A possible way to bypass these issues, suggested in \cite{Aharony:2008wz,Conlon:2018vov,Conlon:2020wmc} in this context, is to examine the problem from a dual perspective through the lens of holography. It is possible, at least in principle, that the consistency or existence of the putative dual CFT may be established more convincingly, since the latter admits a more rigorous (axiomatic) definition.
On the other hand, holographic duals of such scale-separated solutions remain elusive, with putative duals exhibiting a parametrically large gap in the spectrum of single-trace operators. Recently, there has been interest in investigating properties the dual CFT must have, from attempting to formulate a brane picture and studying its properties \cite{Apers:2025pon,Apers:2026lgi, Bedroya:2025ltj}, although full constructions remain challenging to obtain. If one adopts the viewpoint that consistent AdS EFTs must admit a holographic dual, non-trivial constraints should follow. Such constraints have already been formulated in the past in specific settings \cite{Bobev:2023dwx,Perlmutter:2024noo}.

More recently, a new holographic constraint has been proposed \cite{Bobev:2025yxp}. In essence, it states that if the sum of the (tree-level) scaling dimensions of two scalar operators in the putative CFT equals the scaling dimension of a third operator, then the corresponding three-point function must vanish. This can occur easily for theories where the spectrum of conformal dimensions is integer, which was found to be the case for a variety of scale-separated solutions, first in the original papers \cite{Conlon:2021cjk,Apers:2022zjx,Apers:2022tfm} (see also \cite{Herraez:2018vae,Marchesano:2019hfb}) for 4d compactifications and also later in 3d ones \cite{Arboleya:2024vnp,Farakos:2025bwf, VanHemelryck:2025qok}.
Ref.~\cite{Bobev:2025yxp} verified that the DGKT solutions, corresponding to AdS$_4$ vacua from compactifications on the $\mathbb{T}^6/\mathbb{Z}_3\times \mathbb{Z}_3$ orbifold, contain scalars that must be subject to their constraint, and they find that the cubic couplings vanish for the extremal arrangements. Incidentally, this happens for both supersymmetric and non-supersymmetric vacua.

In this paper, we check whether the constraint of ref.~\cite{Bobev:2025yxp} holds for type II solutions on $\mathbb{Z}_2 \times \mathbb{Z}_2 \times \mathbb{Z}_2$ orbifolds leading to the AdS$_3$ vacua of refs.~\cite{VanHemelryck:2025qok,Arboleya:2024vnp,Farakos:2025bwf} that have an integer spectrum. These comprise the scale-separated solution in type IIB on $\mathrm{Nil}_3 \times \mathbb{T}^4$, two solutions on solvmanifolds in type IIB, and a solution on a seven-torus in type IIA. 
We also check the constraint for the scale-separated AdS$_4$ solution on the $\mathbb{T}^6/\mathbb{Z}_2 \times \mathbb{Z}_2$ orbifold, studied explicitely by \cite{Camara:2005dc} and referred to as the CFI construction.
In none of these examples is the constraint satisfied. Surprisingly, we find that in all but one of the cases, the problem is remedied when different orbifold groups are considered. 
In the case of the solvmanifold compactification with only one set of O5-planes of ref.~\cite{VanHemelryck:2025qok} (which is related to some of the non-supersymmetric solutions of ref.~\cite{Arboleya:2024vnp}), the violation of the constraint cannot be resolved by a different orbifold and is therefore ruled out.
For all the other 3d vacua, the curing orbifolds are inevitably non-abelian, and we consider such groups that have the original $\mathbb{Z}_2 \times \mathbb{Z}_2 \times \mathbb{Z}_2$ as a subgroup.
In the case of the scale-separated solution on $\mathrm{Nil}_3 \times \mathbb{T}^4$, we describe the non-abelian orbifold in detail, and we find that we only need two independent O5-planes, wrapping cycles in the same homology class. We discuss the intersection of these O5-planes in the covering space and how this is affected on the orbifold. 
We reach very similar conclusions for the DGKT--CFI construction in 4d on $\mathbb{T}^6/\mathbb{Z}_2 \times \mathbb{Z}_2$ and discuss an extension of the orbifold group to $\mathbb{Z}_4 \times \mathbb{Z}_4$ which also has a $\mathbb{Z}_2 \times \mathbb{Z}_2$ subgroup. 
Remarkably, all these examples seem to suggest that the holographic constraint is only satisfied when O-planes wrap cycles in only one homology class. We discuss how, in such geometries, O-planes that intersect in the covering space should become non-intersecting upon resolving the orbifold singularities \cite{Junghans:2023yue,Montero:2024qtz}.

The main text focuses on the results, while the supporting calculations are included in the appendices. In section~\ref{sec:holo_constraint}, we review the holographic constraint proposed in ref.~\cite{Bobev:2025yxp}, and in section~\ref{sec:orbifolds} we briefly review orbifolds of 6d and 7d twisted tori with only $\mathbb{Z}_2$ factors.
We show in section~\ref{sec:AdS3IIB_cubic_couplings} that the holographic constraint for the scale-separated vacuum of type IIB string theory on $(\mathrm{Nil}_3 \times \mathbb{T}^4)/\mathbb{Z}_2\times\mathbb{Z}_2\times\mathbb{Z}_2$ is not satisfied, and we discuss a new non-abelian orbifold for which it is no longer violated. We also investigate the constraint for other type~II AdS$_3$ vacua in section~\ref{sec:other_AdS3_cubic_couplings}. In section~\ref{sec:AdS4_cubic_couplings}, we show that the constraint is not satisfied for the DGKT--CFI scenario on a $\mathbb{T}^6/\mathbb{Z}_2 \times \mathbb{Z}_2$ orbifold, and we discuss all our results in section~\ref{sec:discussion}.

\section{A holographic constraint on cubic couplings}
\label{sec:holo_constraint}
In this section, we briefly outline the origin and statement of the cubic coupling constraint recently derived in \cite{Bobev:2025yxp}. To do so, let us consider a family of AdS$_{d+1}$ vacua parametrised by a large central charge $c$, in the limit that admits a genuine EFT description. In particular, we assume that such a theory only contains gravity as well as a finite number of fields (with spin $s<2$), and is characterised by a single UV cutoff. This includes, as a particular case, scale-separated vacua, where KK towers can be consistently integrated out. At the two-derivative level, the scalar sector of such a theory is described by the effective Lagrangian (in Euclidean signature)
\begin{equation}
\mathcal{L}_{\mathrm{eff}}=  -R-\frac{d(d-1)}{L_{{\rm AdS}}^2}+\frac{1}{2}\left(\partial_\mu \phi_i\right)^2+\frac{1}{2} \frac{m_i^2}{L_{\rm AdS}^2} \phi_i^2  +d_{i j k} \phi_i \partial_\mu \phi_j \partial^\mu \phi_k+\frac{c_{i j k}}{L_{\rm AdS}^2} \phi_i \phi_j \phi_k+\ldots,
\end{equation}
where the dots denote terms of higher order in the fields $\phi_i$, and repeated indices are summed over. The cubic couplings $c_{ijk}$ and $d_{ijk}$ arise from an expansion of the potential and the kinetic term around the vacuum, respectively. From an AdS/CFT perspective, such vacua define dual, putative holographic CFTs, characterised by a large gap $\Delta_{\rm{gap}} \gg 1$ in the spectrum of single-trace operators. The only single-trace primaries are the operators dual to the scalars $\phi_i$, which in our case will be identified with the moduli of the compactification. According to the usual dictionary, their conformal dimension is determined by
\begin{equation}
    \Delta_i(\Delta_i-d)= m_i^2 L^2_{\rm{AdS}}.
\end{equation}
Of particular importance for us will be \emph{extremal} and \emph{super-extremal} arrangements; triplets of operators $\left\{\mathcal{O}_i\right\}$ whose conformal dimensions add up to each other as
\begin{equation}\label{eq:deltaijk}
\Delta_i+\Delta_j=\Delta_k \qquad \textrm{and} \qquad 
\Delta_i+\Delta_j=\Delta_k -2n \qquad n \in \mathbb{N}^+,
\end{equation}
respectively. While such configurations would appear to be fine-tuned in generic EFTs with a low amount of supersymmetry, they are actually ubiquitous in the classes of (scale-separated) vacua we will examine. This is because such vacua often come with exactly integer spectra at tree level, both in 3 and 4 dimensions \cite{Arboleya:2024vnp,Farakos:2025bwf, VanHemelryck:2025qok,Conlon:2021cjk,Apers:2022zjx,Apers:2022tfm,Quirant:2022fpn,Ning:2022zqx,Plauschinn:2022ztd,Andriot:2023fss,Arboleya:2025ocb,Arboleya:2025jko}.\footnote{Such a property hints at the existence of structure that would be more opaque from a bulk perspective. See \cite{Bonifacio:2018zex,Apers:2022vfp} for a possible interpretation in terms of generalised shift symmetries.}

The holographic constraint on (super-)extremal arrangements can be understood by examining the three-point functions between single-trace primaries. This is unlike usual consistency requirements (such as unitarity) for CFTs, which appear at the level of the spectrum, and rather concerns dynamical aspects of the theory. According to AdS/CFT, the CFT three-point function coefficients can be computed by evaluating the corresponding Witten diagram in the bulk. For three scalar primaries $\left\{\mathcal{O}_i\right\}$, they read 
\begin{equation}\label{eq:Cijk}
C_{i j k}=\eta^{-2} c_{i j k}^{\prime} \frac{\Gamma\left(\Theta_i\right) \Gamma\left(\Theta_j\right) \Gamma\left(\Theta_k\right) \Gamma\left(\Theta-\frac{d}{2}\right)}{2 \pi^d \Gamma\left(\Delta_i-\frac{d}{2}\right) \Gamma\left(\Delta_j-\frac{d}{2}\right) \Gamma\left(\Delta_k-\frac{d}{2}\right)},
\end{equation}
where $\eta=M_\mathrm{Pl}^2/2$ in terms of the $(d+1)$-dimensional Planck mass, and
\begin{equation}
\Theta=\frac{\Delta_i+\Delta_j+\Delta_k}{2} \quad \quad \quad \quad \Theta_n = \Theta-\Delta_n\,.
\end{equation}
The ``effective" cubic coupling $c'_{ijk}$ is defined as
\begin{equation}
c_{i j k}^{\prime}=c_{i j k}+\frac{m_i^2-m_j^2-m_k^2}{2} d_{i j k}.
\end{equation}
Inspecting the arguments of the Gamma functions in \eqref{eq:Cijk}, one can immediately see the appearance of a divergence whenever \eqref{eq:deltaijk} is satisfied, leading to the suspicion that the $c'_{ijk}$ must vanish. This is indeed what happens in theories where the operator dimensions $\Delta_i$ are protected (for example by extended supersymmetry) and the extremality condition \eqref{eq:deltaijk} is exact, as has been known since the early days of holography \cite{DHoker:1999jke} (see also \cite{Aprile:2020uxk} for the connection to a refined notion of single trace operators in the CFT, and \cite{Bobev:2025gzu} for a recent, non-trivial example). In less supersymmetric settings, however, the (unprotected) values of the conformal dimensions will generically receive corrections of the form
\begin{equation}\label{eq:gamma}
\Delta_i=\Delta_i^{(0)}+\frac{\gamma_i}{c}+\ldots,
\end{equation}
where the scaling with $c$ comes from the one-loop correction to the bulk propagator and $\gamma$ is the anomalous dimension. In turn, this will produce an atypical enhancement of the coefficients $C_{ijk}$ by a factor of $c$, for the (super)-extremal arrangements satisfying
\begin{equation}\label{eq:extremal_arrangements}
\Delta_i^{(0)}+\Delta_j^{(0)}=\Delta_k^{(0)} -2n \qquad  n \in \mathbb{N},
\end{equation}
i.e. extremality at zeroth order in a large-$c$ expansion. In ref.~\cite{Bobev:2025yxp}, it was shown how such an enhancement is incompatible with large-$c$ factorisation of the dual CFTs, under the assumptions cited at the beginning of this section. As a consequence, it follows that any consistent, scale-separated EFT with a finite number of fields must obey the cubic coupling constraint
\begin{equation}\label{eq:constraint}
    c'_{ijk}=0\,,
\end{equation}
for any of its extremal arrangements.\footnote{In examples with two notions of a large-$N$-like parameter due to the inclusion of open string modes as in \cite{Chester:2025wti,Chester:2025jxg}, the constraint \eqref{eq:constraint} does not hold in its current form (see also \cite{Castro:2024cmf}). This will not be the case for any of the vacua we examine.} In constructions which are not scale-separated, such as consistent truncations, it is in principle possible for loops of KK modes to enhance the scaling of the anomalous dimension \eqref{eq:gamma} to $1/c^{\alpha}$ with $ \alpha < 1 $, and hence modify the constraint. However, as long as $\alpha > 1/2$, the constraint will still hold in such set-ups.

Remarkably, it was shown in \cite{Bobev:2025yxp} that the constraint is non-trivially satisfied for the simplest incarnation of the DGKT vacua (both for supersymmetric and non-supersymmetric vacua), based on the orbifold $\mathbb{T}^6/\mathbb{Z}_3 \times \mathbb{Z}_3$. The rest of the paper is devoted to a more systematic analysis of the condition \eqref{eq:constraint} in the landscape of scale-separated vacua, in both $3$ and $4$ dimensions. One of our declared goals will be to understand whether the additional structure imposed by \eqref{eq:constraint} can teach us any lessons on the bulk side of the correspondence, such as features of the compactification space.

\section{Comments on orbifolds of twisted tori}\label{sec:orbifolds}
The AdS$_3$ and AdS$_4$ vacua we consider in this paper arise from compactifying type II string theory on a (twisted) torus, orbifolded by a discrete group involving only $\mathbb{Z}_2$ factors. We write such a twisted torus as a non-compact group manifold $\mathcal{M}_n$, quotiented by a discrete lattice group $\Lambda$ consisting of translations:
\begin{equation}
    \tilde{\mathbb{T}}^n = \mathcal{M}_n / \Lambda\,.
\end{equation}
This means that for coordinates $x^m$ on $\mathcal{M}_n$, we identify $x^m = x^m + \Lambda^m$, where $\Lambda^m$ is a lattice vector.
In general, we can write the metric of twisted tori as follows:
\begin{equation}\label{eq:metric_twisted_torus}
    \dd s_n^2 = \sum_{a=1}^n (L_a e^a)^2\,, \qquad e^a = e^a_m \dd x^m\,,
\end{equation}
where the $\{L_i\}$ are radii expressed in \textit{string units}, and the co-frame one-forms $e^a$ satisfy the Maurer-Cartan equations:
\begin{equation}
    \dd e^a = \frac{1}{2}f^{a}{}_{bc}e^b \wedge e^c\,.
\end{equation}
The structure constants $f^a{}_{bc}$ are those of the underlying Lie algebra of the group manifold.
Orbifolds are then obtained by quotienting the space by an orbifold group $\Gamma$. This must be a discrete and finite symmetry group of the lattice $\Lambda$, and its group elements must act affinely on the coordinates as
\begin{equation}
    \Theta(\vec{x}) = \Theta_\circ \cdot \vec{x} + \vec{\Theta}_\sharp\,, 
\end{equation}
where $\vec{\Theta}_\sharp$ is a translation and $\Theta_\circ$ is a rotation that leaves the lattice invariant, i.e. $\Theta_\circ \cdot \vec{\Lambda} = \vec{\Lambda}'$ for two lattice vectors $\vec{\Lambda}$ and $\vec{\Lambda}'$. The group is finite, so every element has a finite order, i.e. there exists a positive integer $n$ for which $\Theta^n(\vec{x}) = \vec{x} + \vec{\Lambda}$. This requires that 
\begin{equation}
\label{eq:orbifold_conditions}
    \Theta_\circ^n( \vec{x}) = \vec{x}\,, \qquad ( 1 + \cdots + \Theta_\circ^{n-1})(\vec{\Theta}_\sharp) = \vec{\Lambda}
\end{equation}
Here,  we are motivated to look at $\mathbb{Z}_2 \times \mathbb{Z}_2 \times \mathbb{Z}_2$ and $\mathbb{Z}_2 \times \mathbb{Z}_2$ orbifolds for the AdS$_3$ and AdS$_4$ vacua of refs.~\cite{VanHemelryck:2025qok,Farakos:2025bwf,Camara:2005dc}. To abbreviate the notation, we write $\mathbb{Z}_2^3\equiv \mathbb{Z}_2 \times \mathbb{Z}_2 \times \mathbb{Z}_2$ and $\mathbb{Z}_2^2 \equiv \mathbb{Z}_2 \times \mathbb{Z}_2$ throughout the whole paper.
Such $\mathbb{Z}_2^3$ orbifolds of (twisted) seven-tori have been discussed extensively for the construction of $G_2$-holonomy manifolds by the seminal work of Joyce \cite{Joyce:1996i,Joyce:1996ii,MR1787733}.
The AdS$_3$ vacua studied later arise as compactifications on a nilmanifold, a flat seven-torus and solvmanifolds. In all cases, we denote the orbifold group to be generated by the elements which we call $\alpha$, $\beta$ and $\gamma$, and we denote the group by $\left<\alpha, \beta, \gamma\right>$. For all these group manifolds, the rotational parts of the orbifolds are given by 
\begin{align}\label{eq:three-Z2_generators}
\begin{split}
    \alpha_\circ ( x^1, x^2, x^3, x^4, x^5, x^6, x^7) &= (-x^1, -x^2, x^3, x^4, -x^5, -x^6, x^7)\\
    \beta_\circ ( x^1, x^2, x^3, x^4, x^5, x^6, x^7) &= (x^1, x^2, -x^3, -x^4, -x^5, -x^6, x^7)\\
    \gamma_\circ ( x^1, x^2, x^3, x^4, x^5, x^6, x^7) &= ( -x^1, x^2, -x^3, x^4, -x^5, x^6, -x^7)\,.
\end{split}
\end{align}
These rotations act on the co-frame one-forms $e^a = e^a_m \dd x^m$ in the same way, so one can take the identities above and replace $x$ by $e$. Note that the Maurer-Cartan equations must be invariant under the orbifold group. Additionally, one sees that there are no one- and two-forms that are invariant under the orbifold group, and that there are only seven three- and four-forms that are invariant:
\begin{gather}
\notag
    \left\{e^{127}\,, e^{347}\,, e^{567}\,, e^{145}\,, e^{136}\,, e^{235}\,,  -e^{246} \right\}\,,\\
\label{eq:basis_Z23_forms}
    \left\{e^{3456}\,, e^{1256}\,, e^{1234}\,, e^{2367}\,, e^{2457}\,, e^{1467}\,, -e^{1357} \right\}\,.
\end{gather}
We used the short notation $e^{a...b} = e^a \wedge ... \wedge e^b$, and we denote the Poincaré dual cycles as $\Pi_{a...b}$. Therefore, we have the Betti numbers $b^1=0$, $b^2=0$ and $b^3=7$. Furthermore, one can equip the orbifold with a $G_2$-structure, and its left-invariant three-form can be written in this basis of three-forms:
\begin{multline}
    \Phi = L_1 L_2 L_7 e^{127}+L_3 L_4 L_7 e^{347}+L_5 L_6 L_7 e^{567}\\
    + L_1 L_4 L_5 e^{145}+ L_1 L_3 L_6 e^{136}+L_2 L_3 L_5 e^{235} - L_2 L_4 L_6 e^{246}\,.
\end{multline}
For the $\mathbb{Z}_2^2$ orbifolds that we consider for the four-dimensional AdS vacua in this paper, we define $\alpha$ and $\beta$ to be the two $\mathbb{Z}_2$ involutions that generate the group, for which their rotational parts act on the 6d coordinates in a very similar way to the seven-dimensional case:
\begin{align}\label{eq:two-Z2_generators}
    \alpha_\circ ( x^1, x^2, x^3, x^4, x^5, x^6) &= (-x^1, -x^2, x^3, x^4, -x^5, -x^6)\\
    \beta_\circ ( x^1, x^2, x^3, x^4, x^5, x^6) &= (x^1, x^2, -x^3, -x^4, -x^5, -x^6)\,,
\end{align}
and similar for the co-frame one-forms.
There are no one-forms that remain invariant under the orbifold, and the only invariant two- and three-forms are
\begin{gather}
    \left\{e^{12}\,, e^{34}\,, e^{56} \right\}\,, \quad 
    \left\{e^{145}\,, e^{136}\,, e^{235}\,, -e^{246}\, ;\, e^{236}\,, e^{245}\,, e^{146}\,, -e^{135} \right\}\,.
\end{gather}
This leads to an orbifold with Hodge numbers $h^{1,1}=3$ and $h^{2,1}=3$, corresponding to three K\"ahler moduli and three complex structure moduli. Note that all of these scalars can be expressed in terms of the six radii.

\section{Cubic couplings for scale-separated \texorpdfstring{AdS$_3$}{AdS3} solutions in type IIB string theory}\label{sec:AdS3IIB_cubic_couplings}
We want to check the constraint~\eqref{eq:constraint} for the recently found scale-separated solution in type IIB string theory \cite{VanHemelryck:2025qok}, whose mass spectrum leads to integer scaling dimensions for dual operators in the putative dual field theory. 
The AdS$_3$ solution arises as a compactification on a $\mathbb{Z}_2^3$ orbifold of $\mathrm{Nil}_3 \times \mathbb{T}^4$, where the first factor refers to the 3d nilmanifold, and the whole 7d manifold has $G_2$-structure. This space is also called $\mathfrak{n}_2$ in the mathematics literature \cite{MR2698220,MR2811660,MR3739330,MR4626831,MR4789076}, and is characterised by the following Maurer-Cartan equations:
\begin{equation}\label{eq:MC_Nil3_T4}
    \dd e^{7} = \omega\; e^{12}\,, \qquad \dd e^{1,2,3,4,5,6}=0\,,
\end{equation}
such that directions (1,2,7) parametrise the nilmanifold and (3,4,5,6) the four-torus. The setup requires $F_3$- and $F_7$-flux, which in the basis \eqref{eq:basis_Z23_forms} take the following form:
\begin{equation}\label{eq:F3-flux}
    F_{3} = F_{127}e^{127}- F_{347}e^{347}-F_{567}e^{567}+F_{145}e^{145}+F_{136}e^{136}+F_{235}e^{235}-F_{246}e^{246}\,, \quad F_{7}=f_7 e^{1234567}\,,
\end{equation}
where the $\{F_{ijk}\}$ and $f_7$ are quantised flux parameters.
Two sets of O5-planes are necessary for tadpole cancellation, along the cycles $\Pi_{347}$ and $\Pi_{567}$. All these ingredients stabilise the eight scalars $L_1,... L_7$ and $g_s$, and we note that all axions are projected out by the $\mathbb{Z}_2^3$ orbifold. The details about the solution were worked out in ref.~\cite{VanHemelryck:2025qok} and are reviewed in appendix~\ref{app:AdS3_Nil3_T4}. The solutions are obtained by solving the supersymmetry equations in terms of bispinors, a procedure studied before in refs.~\cite{Dibitetto:2018ftj,Passias:2019rga,Passias:2020ubv, Emelin:2021gzx,VanHemelryck:2022ynr}. Importantly, the solution becomes scale-separated in the well-controlled regime after taking a large flux-limit.
We then diagonalise the mass matrix in the vacuum and denote the mass eigenstates as $\xi$, $\{\chi_{1,2,3}\}$ and $\{\sigma_{3,4,5,6}\}$. The transformations between these mass eigenstates and the scalars $L_1,...L_7$ and $g_s$ are included in appendix~\ref{app:AdS3_Nil3_T4}. Moreover, the mass spectrum of the scalars is such that the conformal dimensions of the putative dual operators are integers, with the following values:
\begin{equation}
    \Delta( \xi ) = 12\,, \qquad \Delta(\chi_{1,2,3}) = 4\,, \qquad \Delta(\sigma_{3,4,5,6}) = 4\,.
\end{equation}
The spectrum does not allow for extremal arrangements, but super-extremal ones of the type (12,4,4) do appear.
The holographic constraint \eqref{eq:constraint} informs us that the associated cubic couplings $c'_{ijk}$ should vanish. There are no derivative couplings present, as the fields involved are all saxions, and there are no axions. This means that only the cubic coupling coming from the potential contributes. If we expand the potential around the vacuum in terms of the mass eigenstates and evaluate the cubic contributions involving the above super-extremal arrangements (12,4,4), we find that such cubic couplings do not vanish, as 
\begin{equation}\label{eq:cubic_couplings_Nil3_T4}
    L_\mathrm{AdS}^2 V_\mathrm{cubic} \supset \frac{1}{3!}162 \sqrt{\frac{2}{7}} \;\tilde{\xi} \; (\tilde \sigma_3+ \tilde \sigma_4 - \tilde \sigma_5 -\tilde \sigma_6)^2\,,
\end{equation}
where tilded scalars parametrise fluctuations away from the vacuum expectation values of the fields, denoted by an asterisk, e.g. $\tilde \xi = \xi - \xi^*$. The $\{\sigma_{3,4,5,6}\}$ are the canonically normalised radii $\sigma_{3,4,5,6} = 2^{-1/2} \log(L_{3,4,5,6})$, and the scalar $\xi$ can be expressed in terms of the lower-dimensional dilaton and the radii $L_1$, $L_2$ and $L_7$, as given in appendix~\ref{app:AdS3_Nil3_T4}. Since the contribution~\eqref{eq:cubic_couplings_Nil3_T4} does not vanish, the holographic constraint \eqref{eq:constraint} is not satisfied and therefore invalidates this solution. However, we see that if we impose
\begin{equation}
    \sigma_3 + \sigma_4 = \sigma_5 + \sigma_6 \quad \Leftrightarrow \quad L_3 L_4 = L_5 L_6\,,
\end{equation}
the offending cubic coupling vanishes. This suggests that the constraint~\eqref{eq:constraint} can still be satisfied for this background if we consider a different orbifold than $\mathbb{Z}_2^3$ that would enforce this.

In fact, we remark also that the cubic couplings \eqref{eq:cubic_couplings_Nil3_T4} do only arise due to very specific terms in the potential: those terms generated by the two sets of O5-planes wrapping cycles \textit{in distinct homology classes} and the three-form fluxes $F_{347}$ and $F_{567}$ participating in the Bianchi identity for $F_3$, and hence constrained by tadpole cancellation with the O5-planes. Therefore, the holographic constraint hints at something being amiss in this O5-plane configuration.
Indeed, curing it with an orbifold that sets $L_3 L_4 = L_5 L_6$, will lead to O5-planes wrapping cycles \textit{in only one homology class}. We discuss such an orbifold in the next section.

\subsection{A non-abelian orbifold to the rescue}
Instead of taking the classic $\mathbb{Z}_2^3$ orbifold above, we look at the orbifold group $\left< \alpha,\tau,\gamma\right>$ generated by the following elements:
\begin{align}
    \alpha ( x^1, x^2, x^3, x^4, x^5, x^6, x^7) &= \left(-(x^1+1), -x^2, x^3+1, x^4, -x^5+1, -x^6, x_1^7 \right)\\
    \tau ( x^1, x^2, x^3, x^4, x^5, x^6, x^7) &= (-x^1, -x^2, x^5, x^6, x^3, x^4, x^7 +1)\\
    \gamma ( x^1, x^2, x^3, x^4, x^5, x^6, x^7) &= ( -(x^1+1), x^2+1, -x^3, x^4+1, -x^5, x^6+1, -x_{1,2}^7)\,.
\end{align}
Note that we are orbifolding a nilmanifold where all coordinates have a periodicity of 2, and for which half lattice vector shifts in $x^1$ and $x^2$ require additional shifts in $x^7$, hence we defined $x^7_{1} \equiv x^7 -x^2/2$ and $x^7_{1,2} \equiv x^7 + x^1/2 -x^2/2$. We comment on this more in appendix~\ref{app:shifts_nil}.
Compared to the original orbifold, we have replaced the rotation $\beta_\circ$ with $\tau_\circ$, which swaps two pairs of coordinates, $(x^3,x^4) \leftrightarrow (x^5,x^6)$.
The important properties of this group are that all generators are of order 2, i.e. $\alpha^2 = \tau^2 = \gamma^2 =1$ and  $\gamma$ commutes with both $\alpha$ and $\tau$. The elements $\alpha$ and $\tau$ do not commute, but they generate the group $\left< \alpha, \tau \right>$ of order 8, which is isomorphic to the dihedral group $D_4$. It has the elements
\begin{equation}
    \left<\alpha ,\tau \right> = \{1\,, \alpha\,, \tau\,, \alpha \tau \,, \tau \alpha\,, \alpha \tau \alpha\,, \tau \alpha \tau\,, \alpha \tau \alpha \tau = \tau \alpha \tau \alpha\}
\end{equation}
One appealing aspect of the orbifold group is that
\begin{equation}
    \tau  \alpha  \tau  \alpha = \alpha  \tau \alpha \tau = \beta\,,
\end{equation}
with $\beta$ acting as in eq.~\eqref{eq:three-Z2_generators}, which means that $\left< \alpha, \beta, \gamma \right>$ is a $\mathbb{Z}_2^3$ subgroup of $\left< \alpha, \tau, \gamma \right>$.
This means that the cohomology of the orbifold $\left<\alpha, \tau,\gamma\right>$ is contained into that of $\left<\alpha,\beta,\gamma\right>$.
The reason that we introduced this orbifold is that it requires
\begin{equation}
    L_3 = L_5\,, \qquad L_4 = L_6\,.
\end{equation}
which can be seen from the fact that the metric \eqref{eq:metric_twisted_torus} has to be invariant under the orbifold. Therefore, it projects out the scalar field combination that led to the violation of the cubic coupling constraint above.
The cohomology changes appropriately, as
\begin{equation}
    \tau(e^{347}+e^{567}) = e^{347}+e^{567}\,, \qquad \tau(e^{145} + e^{136}) = e^{145} + e^{136}\,,
\end{equation}
but which is not valid for the terms separately, as $\tau(e^{347}) = e^{567}$ and $\tau(e^{145})= e^{136}$. As a consequence, there are only 5 invariant three-forms, and the left-invariant three-form $\Phi$ takes the form 
\begin{align}\notag
    \Phi =& \;L_1 L_2 L_7 e^{127}+ L_3 L_4 L_7 \left( e^{347}+ e^{567}\right)
    + L_1 L_{3} L_{4} \left( e^{145}+ e^{136}\right)\\
    &+L_2 L_{3}^2 e^{235} - L_2 L_{4}^2 e^{246}\,.
\end{align}
The shift vectors in the orbifold generators make the singularity structure of the orbifold rather simple, and are such that an additional orientifold involution only generates the required O-planes. First, let us focus on the orbifold singularities. There is only one group element that has fixed points, namely $\alpha \tau \alpha \tau =  \beta$, and they are localised only in the $\mathbb{T}^4$, at
\begin{equation}
    x^{3,4,5,6} = \{0,1\}.
\end{equation}
Of these $2^4=16$ fixed points, $\alpha$ swaps one half by the other half, reducing it to 8 distinct fixed points, and $\gamma$ identifies a further half of it, while $\tau$ does not identify any of the remaining ones, leading to only 4 sets of independent fixed points. These orbits of fixed points have the following representatives
\begin{equation}
   (x^3, x^4, x^5, x^6)\; : \qquad  (0,0,0,0)\,, \quad (1,0,0,0)\,, \quad (0,1,0,0)\,, \quad (1,1,0,0)\,.
\end{equation}
All these fixed points introduce singularities of the type $\mathbb{C}^2/\mathbb{Z}_2$, which can be resolved by the standard blow-up procedure with extra blow-up moduli, which is work that we carry out in the future.

Next, we impose an O5-involution, which we take now to be the same as the rotational part of $\tau$,
\begin{equation}
    \sigma_\tau ( x^1, x^2, x^3, x^4, x^5, x^6, x^7) = \left(-x^1, -x^2, x^5, x^6, x^3, x^4, x^7\right)\,.
\end{equation}
Only the elements $\sigma_\tau$ and $\sigma_\tau \alpha \tau \alpha \tau  = \sigma_\tau \beta$ have fixed points, leading to O5-planes. For this, it is useful to introduce the coordinates
\begin{equation}
    u_{\pm} = x^3 \pm x^5\,, \qquad v_{\pm} = x^4 \pm x^6\,.
\end{equation}
In terms of these coordinates, the orientifold fixed points lie at
\begin{align}
    \sigma_\tau\;:&  \quad x^{1,2} = \left\{0, 1 \right\}\,, \quad u_+ = v_+ = 0\,, \\
    \sigma_\tau \beta\;:&  \quad  x^{1,2} = \left\{0, 1\right\}\,, \quad u_- = v_- = 0\,. 
\end{align}
Also here, the number of independent fixed points is reduced due to the orbifold. For instance, we see that $\gamma$ identifies half of the images within each set, as it identifies the fixed points $x^{1,2}=0 \leftrightarrow x^{1,2} =1$. We can parametrise this by  the following representatives:
\begin{gather}
    (x^{1},x^{2},u_+, v_+): \qquad \left(0,0, 0, 0\right)\,, \qquad \left(1,0, 0, 0\right)\\
    (x^{1},x^{2},u_-, v_-): \qquad \left(0,0, 0, 0\right)\,, \qquad \left(1,0, 0, 0\right)
\end{gather}
The generator $\tau$ does not further identify any of the O5-planes. 
Importantly, $\alpha$ identifies all the fixed points of $\sigma_\tau$ with those of $\sigma_\tau \beta$, as it interchanges $u_\pm \leftrightarrow v_\pm$. It also follows from the fact that $\sigma_\tau$ and $\sigma_\tau \beta$ belong to the same conjugacy class. Indeed, a consistent orientifold involution $\sigma$ must satisfy \cite{Gimon:1996rq}\footnote{Note that in \cite{DeWolfe:2005uu}, the orbifold generators and orientifold involution do not satisfy this condition. This issue was raised in ref.~\cite{Junghans:2023yue}, and remedied by imposing a different orientifold involution.}
\begin{equation}\label{eq:invc}
    g^{-1} \sigma g = \sigma g' + \Lambda\,, \quad \sigma g^{-1} \sigma g = g' + \Lambda\,,
\end{equation}
where $g$ and $g'$ are orbifold group elements.
For the abelian $\mathbb{Z}_2^3$ orbifolds with the typical orientifold involutions used in the literature, $g'$ is the identity, but this is not always the case here, as we have
\begin{equation}
    \alpha \sigma_\tau \alpha = \sigma_\tau \beta\,.
\end{equation}
In the end, there are only two distinct O5-planes that are not mapped to each other under the orbifold elements, and can be parametrised by the following representatives:
\begin{equation}
    (x^{1},x^{2},u_+, v_+) = \left(0,0, 0, 0\right)\,, \qquad (x^{1},x^{2},u_+, v_+) = \left(1,0, 0, 0\right)\,,
\end{equation}
or alternatively,
\begin{equation}
    (x^{1},x^{2},u_+, v_+)= \left(0,0, 0, 0\right)\,,\qquad (x^{1},x^{2},u_-, v_-)= \left(0,0, 0, 0\right)\,.
\end{equation}
It is important to note that, from the perspective of the covering space, these two representatives of the O5-planes intersect at $u_+=u_-=v_+=v_-=0$, which is precisely at an orbifold singularity of the form $\mathbb{C}^2/\mathbb{Z}_2$ or $A_1$ in the ADE classification. From the standard blow-up procedure of this singularity, one can derive that the two O5-planes do not intersect transversally on the exceptional divisor but overlap, similar to the results of ref.~\cite{Junghans:2023yue}. Indeed, one can define the complex coordinates 
\begin{equation}
    z_1 = v_+ + i u_-\,, \qquad z_2 = u_+ + i v_-\,,
\end{equation}
for which the O-plane loci are given by $\Re(z_1)=0=\Re(z_2)$ and $\Im(z_1) = 0= \Im(z_2)$. We can then parametrise the $A_1$ singularity as
\begin{equation}\label{eq:A1_polynomial}
    xy - z^2 =0\,,
\end{equation}
with $x = z_1^2$, $y=z_2^2$ and $z= z_1 z_2$. 
For the blow-up, we take the homogeneous coordinates $x = t X$, $y= t Y$ and $z = t Z$, for which the polynomial \eqref{eq:A1_polynomial} becomes $t^2(XY-Z^2)=0$, and the exceptional divisor $E$ is given by $t=0$. Working in the affine chart $X \neq 0$ (so that $z_1 \neq 0$) and setting $X=1$, we obtain $x = t$ and hence $Y= y/x = z_2^2/z_1^2$. Points on the exceptional divisor correspond to approaching the origin with fixed ratio $z_2/z_1$, so that $Y$ remains finite as $t \to 0$. Parametrising $z_1 = a_1 + i b_1$ and $z_2=a_2 +  i b_2$, we notice that for the O5-plane at $\Re(z_1)=0=\Re(z_2)$, we have $Y= b_2^2/b_1^2 \in \mathbb{R}_+$, while for the O5-plane $\Im(z_1)=0=\Im(z_2)$ we have $Y =a_2^2/a_1^2\in \mathbb{R}_+$. Thus both loci intersect the exceptional divisor along the same real subset $Y \in \mathbb{R}_+$. One concludes that the O5-planes are not transversally intersecting on the exceptional divisor, but coincide on $E$. 

Finally, the O5-planes source the $F_3$ Bianchi identity. From the perspective of the covering space, the O5 charge distribution is written as a sum of Dirac-delta distributions as follows:
\begin{equation}
    j_\mathrm{O5} \propto\sum_{x^1_*,  x^2_*\in \{0,1\}} \delta(x^1-x^1_*)\delta(x^2-x^2_*)\dd x^1 \wedge \dd x^2 \wedge \left[\delta(u_+)\delta(v_+)\dd u_+ \wedge \dd v_+ + \delta(u_-)\delta(v_-)\dd u_- \wedge \dd v_-\right]\,.
\end{equation}
Notice that smearing, which replaces the Dirac-Delta distribution by a constant distribution, would indeed reduce this to 
\begin{equation}
    j_\mathrm{O5}^\mathrm{smear} \propto\dd x^1 \wedge \dd x^2 \wedge \left(\dd u_+ \wedge \dd v_+ + \dd u_- \wedge \dd v_-\right) = 2 (e^{1234} +  e^{1256})\,,
\end{equation}
which is indeed the four-form that $\dd F_3$ is proportional to in the smeared approximation.

Recently, ref.~\cite{Tringas:2025bwe} advocated for a similar scale-separated construction in heterotic string theory. There, the internal manifold and orbifold are the same as here, and it can be argued that the construction can be obtained by S-duality from the solution of \cite{VanHemelryck:2025qok}. Therefore, we expect that the holographic constraint will not be satisfied for the solution of ref.~\cite{Tringas:2025bwe} but can be remedied in the same way as explained above. However, it remains unclear what this would imply geometrically for the gravitational instantons that are necessary for tadpole cancellation in that setting.

\section{Cubic couplings for other \texorpdfstring{AdS$_3$}{AdS3} vacua}\label{sec:other_AdS3_cubic_couplings}
The non-vanishing of cubic couplings for (super-)extremal configurations of scalar operators is not specific to the scale-separated solution discussed above. Indeed, other AdS$_3$ solutions of type II string theory with integer spectra, allowing for such configurations, also violate the constraint. These include the type IIB solutions on solvmanifolds of ref.~\cite{VanHemelryck:2025qok} (see also \cite{Arboleya:2024vnp}) and the massive type IIA solution of ref.~\cite{Farakos:2025bwf}, all arising from $\mathbb{Z}_2^3$ orbifolds of (twisted) tori.
We illustrate this below and argue that, once again, the situation can be cured by considering suitable non-abelian orbifolds. We will not, however, discuss these orbifolds and their associated orientifolds in the same level of detail as above, but some of their aspects are discussed in appendix~\ref{app:other_orbifolds}.

\subsection{Type IIB \texorpdfstring{AdS$_3$}{AdS3} compactifications using solvmanifolds}
In ref.~\cite{VanHemelryck:2025qok}, AdS$_3$ solutions were also explored in the context of compactifications on solvmanifolds. Two setups were considered, one with four sets of O-planes extended along different three-cycles, and another one with just one set of O-planes along one three-cycle. The latter is related to one of the non-supersymmetric solutions found in ref.~\cite{Arboleya:2024vnp,Arboleya:2025ocb}. 
Both setups consider a solvmanifold, which is characterised by the following Maurer-Cartan equations
\begin{equation}\label{eq:MC_Solv7}
    \dd e^{i} = \omega_i e^{i+1} \wedge e^7\,, \quad \dd e^{i+1} = - \omega_i e^{i} \wedge e^7\,, \qquad \dd e^7 = 0\,, \qquad i = 1,3,5\,,
\end{equation}
and which one can orbifold with a $\mathbb{Z}_2^3$ group. Also, this manifold can be equipped with a $G_2$-structure as in section~\ref{sec:orbifolds}. These two setups have the feature that all cycles become parametrically small in AdS units by a large flux limit, but they do not achieve scale separation in the strict sense that the lowest KK mass of a scalar, estimated by the lowest non-trivial eigenvalue of the scalar Laplacian, does not decouple from the AdS scale. In this sense, the compactifications do not describe AdS EFTs but consistent truncations at best. Although it is non-trivial to estimate the large-$c$ scaling of the anomalous dimensions in this scenario, as explained in section~\ref{sec:holo_constraint}, we will assume below that the solutions are subject to the constraint.

\subsubsection*{Solution requiring four sets of O5-planes}
The solutions with four sets of O5-planes arise from taking a specific flux arrangement, which in the basis \eqref{eq:basis_Z23_forms} can be written as:
\begin{equation}\label{eq:RR_Solv7_4O5}
    F_3 = f_{3,1} e^{127} + f_{3,3}e^{347} +  f_{3,5}e^{567} - \tilde{f_3}\left(e^{145}+e^{136}+e^{235}-e^{246}\right)\,, \quad F_7 = -f_7  e^{1234567}\,.
\end{equation}
The parameters $f_{3,i}$, $\tilde f_3$ and $f_7$ are quantised flux parameters.
The four sets of O5-planes wrap the cycles $\Pi_{145}$, $\Pi_{136}$, $\Pi_{235}$ and $\Pi_{246}$.
All moduli are stabilised, and we denote the eight mass eigenstates as $\zeta$, $\{\chi_{1,2,3,4}\}$ and $\{\rho_{1,3,5}\}$. The masses lead to the following scaling dimensions for the dual operators:
\begin{equation}
    \Delta(\zeta) = 3\,, \qquad \Delta(\chi_i) = 4\,, \qquad \Delta(\rho_j) =  1 + 6\left|\frac{\omega_i}{\omega_\Sigma}\right|\,, \qquad i = 1,...4\,,\quad  j = 1,3,5\,,
\end{equation}
where $\omega_\Sigma \equiv \omega_1 + \omega_3+\omega_5$. 
We see that no (super-)extremal arrangements arise amongst the operators dual to $\xi$ and the $\{\chi_i\}$, but such arrangements can arise when at least one of the operators dual to the $\{\rho_j\}$ is involved, depending on the values of the structure constants $\omega_{1,3,5}$. In appendix \ref{app:AdS3_Solv7_4O5}, we compute the cubic couplings of those scalars, and they do not vanish automatically: they depend on the structure constants $\omega_{1,3,5}$. Imposing the cubic coupling constraint \eqref{eq:constraint} therefore leads to a constraint on the possible values of $\omega_{1,3,5}$. Notably, no issues arise if the scalars ${\rho_j}$ are projected out of the spectrum, which can be realised by considering a larger orbifold group. Indeed, the scalars ${\rho_j}$ are related to the radii as follows:
\begin{equation}
    \e^{2\rho_1} = \frac{L_2}{L_1}\,, \qquad \e^{2\rho_3} = \frac{L_4}{L_3}\,, \qquad \e^{2\rho_5} = \frac{L_6}{L_5}\,.
\end{equation}
So using an orbifold which sets $L_1=L_2$, $L_3=L_4$ and $L_5 = L_6$, would indeed project these scalars out.

Furthermore, we remark also here that the non-vanishing cubic couplings arise completely from the terms in the potential that involve the tadpole ingredients: those terms generated by the flux parameters $\tilde{f}_3$ and the O5-planes wrapping three distinct cycles. 
In appendix \ref{app:orbifold_Solv7}, we do consider an orbifold that sets $L_1=L_2$, $L_3=L_4$ and $L_5 = L_6$, which results again in O5-planes wrapping cycles in the same homology class.

\subsubsection*{Solution requiring one set of O5-planes}
For the solution with one set of O5-planes, the structure constants are related as $-\omega_1 = \omega_3 = \omega_5$ and also the three-form fluxes are different and must be taken as follows:
\begin{equation}\label{eq:RR_Solv7_1O5}
    F_3 = -\left(f_{3,1} e^{127} + f_{3,3}e^{347} +  f_{3,5}e^{567}\right) + \tilde{f_3}\left(e^{145}+e^{136}-e^{235}-2e^{246}\right)\,, \quad F_7 = f_7  e^{1234567}\,,
\end{equation}
where again the $\{f_{3,i}\}$, $\tilde f_3$ and $f_7$ are quantised flux parameters.
We need one set of O5-planes, extended along the cycle $\Pi_{246}$, for tadpole cancellation. More details can be found in appendix~\ref{app:AdS3_Solv7_1O5}.
Also here we achieve moduli stabilisation, and we denote the mass eigenstates as $\{\xi_{1,2,3}\}$, $\{\chi_{1,2,3,4}\}$ and $\theta$, for which the dual operators have the following dimensions:
\begin{equation}
    \Delta(\xi_i) = 8\,, \qquad \Delta(\chi_i) = 4\,, \qquad \Delta(\theta)=2\,.
\end{equation}
The potentially dangerous extremal arrangements are given by (4,2,2) and (8,4,4), with corresponding cubic couplings of the schematic form $\theta^2 \chi$ and $\chi \chi \xi$. These all vanish. The cubic couplings $\xi \chi \theta$ corresponding to the super-extremal arrangement (8,4,2) also vanish.

However, the cubic couplings of the schematic form $\xi\theta^2$, corresponding to the super-extremal arrangements $(8,2,2)$, do not all vanish. Indeed, we see that
\begin{equation}
    L_\mathrm{AdS}^2 V_\mathrm{cubic} \supset \frac{1}{3!} 24 \sqrt{30}\;\tilde{\xi}_3 \;\tilde{\theta}^2\,,
\end{equation}
where we defined $\tilde \xi_3 = \xi_3- \xi_{3,*}$ and $\tilde{\theta}=\theta-\theta_*$ to be the fluctuations away from the vacuum, denoted by the asterisks. Also here, the terms in the potential generated by the tadpole ingredients are completely responsible for this cubic coupling.
Contrary to the previous examples, we cannot consider another orbifold to project either $\xi_3$ or $\theta$ out, as in terms of the radii and the dilaton, they are
\begin{equation}
    \e^{4\frac{\sqrt{10}}{3} \;\xi_3} = \delta_3^2 L_7^3 \left(\frac{L_1 L_3 L_7}{L_2 L_4 L_6}\right)^{5/3}\,, \qquad \e^{\sqrt{8}\; \theta} =  \delta_3^2 L_7^3 \frac{L_2 L_4 L_6}{L_1 L_3 L_5} = \left(g_s \frac{L_7}{L_1 L_3 L_5}\right)^2\,.
\end{equation}
Because of the dilaton dependence, any larger orbifold cannot project these scalars out.
Therefore, we conclude that this model is ruled out by the holographic constraint \eqref{eq:constraint}.

\subsection{Massive type IIA \texorpdfstring{AdS$_3$}{AdS3} compactifications on (co-)closed \texorpdfstring{$G_2$}{G2}-structures}\label{sec:mIIA_AdS3_T7}
Finally, we end this section discussing the AdS$_3$ solution constructed in ref.~\cite{Farakos:2025bwf} that also exhibits integer conformal dimensions.
This is a compactification of massive type IIA on a closed and co-closed $G_2$-structure, realised on a $\mathbb{Z}_2^3$ orbifold of the flat torus $\mathbb{T}^7$. The construction consists of $F_0$-flux, four $F_4$-fluxes, and three $H$-fluxes, which in the basis \eqref{eq:basis_Z23_forms} can be parametrised as follows:
\begin{equation}\label{eq:T7_fluxes}
        H = h_1 e^{127}+h_3 e^{347}+h_5 e^{567}\,, \quad F_4 = f_{41} e^{2367}+f_{42}e^{2457}+f_{43}e^{1467}-f_{44}e^{1357}\,, \quad  F_0\,,
\end{equation}
and both O2/D2-sources and O6-planes are required for tadpole cancellation, the O6-planes being extended along the four-cycles $\Pi_{1234}$, $\Pi_{3456}$ and $\Pi_{1256}$.
In this case, all moduli can be stabilised except for one, which we will denote by $\theta_3$. We denote the mass eigenstates as $\xi$, $\{\chi_{1,2,3,4}\}$ and $\{\theta_{1,2,3}\}$, and their putative dual operators have the following dimensions:
\begin{equation}
    \Delta( \xi) = 8\,, \qquad \Delta(\chi_i) = 4\,, \qquad \Delta(\theta_j) = 2\,.
\end{equation}
Although we have three massless moduli $\{\theta_{1,2,3}\}$, the first two are actually stabilised, as in \cite{Rajaguru:2024emw,Becker:2024ijy,Becker:2024ayh}, whereas $\theta_3$ is a true modulus.
Just as in all other examples, it turns out that all extremal arrangements have no cubic couplings. However, there are cubic couplings corresponding to super-extremal arrangements (8,2,2), i.e. of the form $\theta \theta \xi$.
They contribute to the potential as follows: 
\begin{equation}
    L_\mathrm{AdS}^2 V_\mathrm{cubic} \supset -\frac{1}{3!} 16 \sqrt{30} \; \tilde \xi \; (\tilde \theta_1^2 + \tilde \theta_2^2)\,.
\end{equation}
Also here, only the terms in the potential involving the ingredients participating in the tadpole (i.e. $H$-flux and the O6-planes) are responsible for these cubic couplings.
When we express the scalars $\theta_1$ and $\theta_2$ in terms of the radii, we find that
\begin{equation}
    \e^{\sqrt{24}\;\theta_1} = \frac{L_3 L_4 L_5 L_6}{L_1^2 L_2^2}\,, \qquad \e^{\sqrt{8}\;\theta_2} = \frac{L_5 L_6}{L_3 L_4}\,.
\end{equation}
Again we can consider a new orbifold that enforces $L_1 L_2 = L_3 L_4 = L_5 L_6$, such that the scalars $\theta_1$ and $\theta_2$ are projected out, leading to no super-extremal couplings. In appendix~\ref{app:orbifold_T7}, we consider an orbifold group that is isomorphic to $A_4 \times \mathbb{Z}_2$ and which projects the required scalars out.

\section{Cubic couplings for scale-separated \texorpdfstring{AdS$_4$}{AdS4} in type IIA string theory}\label{sec:AdS4_cubic_couplings}

Perhaps the best known class of examples of (classical) scale-separated vacua is the AdS$_4$ DGKT--CFI construction \cite{DeWolfe:2005uu,Camara:2005dc}, obtained by compactifying massive type IIA string theory on a Calabi-Yau orientifold with O$6$-planes. The AdS$_3$ vacua of the previous section share many features with them, including the presence of flux parameters, here $e_i$, that are unconstrained by tadpole considerations and can be tuned to infinity, as to realise scale separation parametrically. The study of their putative dual CFT has attracted significant attention throughout the years \cite{Aharony:2008wz,Conlon:2021cjk,Apers:2022tfm,Apers:2022vfp,Bobev:2023dwx,Montero:2024qtz,Apers:2025pon,Apers:2026lgi, Bedroya:2025ltj}, uncovering a number of puzzling features. Two notable examples are an unusually fast scaling of their central charge, and the anomalous behaviour of certain terms in the CFT partition function \cite{Bobev:2023dwx}. More importantly for our purposes, it was also shown in \cite{Conlon:2021cjk,Apers:2022tfm} that the conformal dimensions for the low-lying operators dual to the scalars of the compactification are always integers. Even more surprisingly, the latter take universal values for \emph{arbitrary} compactification manifolds (orientifoldable CYs), which do not depend on any microscopic details such as fluxes, triple intersection numbers, etc. For supersymmetric vacua, such a universal spectrum is given by
\begin{equation}\label{eq:deltaK}
    \Delta_1= 10 \quad \quad \Delta_2= ...=\Delta_{h_-^{1,1}+1}=6 \quad \quad \tilde{\Delta}_1= 11 \quad \quad \tilde{\Delta}_2= ...=\tilde{\Delta}_{h_-^{1,1}+1}=5
\end{equation}
for the sector including all the K\"ahler moduli and the dilaton ($\Delta$), and the corresponding axions $(\tilde{\Delta})$. In the complex structure sector, it is instead given by
\begin{equation}\label{eq:deltaCS}
    \Delta_1=...=\Delta_{h_-^{2,1}}=2 \quad \quad \quad \tilde{\Delta}_1=...=\tilde{\Delta}_{h_-^{2,1}}=3.
\end{equation}
A striking feature of \eqref{eq:deltaCS} is the appearance of $h^{2,1}$ flat directions, corresponding to marginal deformations of the dual CFT. In this minimally supersymmetric setup, we expect that non-perturbative effects (such as D2-brane instantons) will generate corrections that lift these unprotected flat directions. We aim to show that pathologies do indeed arise when the scalars remain exactly marginal when such effects are ignored.
Notice how, in both cases, the structure of 3d $\mathcal{N}=1$ CFT multiplets forces $|\Delta-\tilde{\Delta}|=1$ \cite{Cordova:2016emh}. It is evident how the spectrum \eqref{eq:deltaK}-\eqref{eq:deltaCS} admits a large number of extremal arrangements, leading to potential constraints on the validity of such vacua. Remarkably, it was recently shown that the cubic coupling constraints are all satisfied in the simplest incarnation of DGKT based on the toroidal orbifold $\mathbb{T}^6/\mathbb{Z}_3^2$. This is true both for the supersymmetric and non-supersymmetric vacua, and results from non-trivial cancellations.\footnote{The non-supersymmetric vacua have fully integer spectra, that differ from \eqref{eq:deltaK} and \eqref{eq:deltaCS}.} The $\mathbb{T}^6/\mathbb{Z}_3^2$ model, however, has some very particular features: there are no geometric complex structure moduli, i.e. $h^{2,1}=0$. From the perspective of the covering space, the O6-plane images self-intersect, but wrap a single cycle in the compactification manifold. Moreover, it was shown in \cite{Junghans:2023yue} that the self-intersections are removed by blowing up the orbifold singularities. The purpose of this section is to explore precisely whether the cubic coupling constraint is still satisfied in models with intersecting O-planes wrapping distinct 3-cycles, especially in the $\mathbb{Z}_2^2$ orbifold, and what can be done if this is not the case.

We consider massive type IIA vacua, compactified on an orientifold of $\mathbb{T}^6 / \mathbb{Z}_2^2$ as in the CFI construction \cite{Camara:2005dc}, analogous to DGKT. The orbifold group is generated by the following $\mathbb{Z}_2$-generators, acting on the complex coordinates $\{z_{1,2,3}\}$ as:
\begin{equation}\label{eq:Z2xZ2}
    \alpha\left(z_1, z_2, z_3\right) = \left(- z_1,  z_2, - z_3\right),  \qquad
     \beta\left(z_1, z_2, z_3\right) =\left( z_1,  -z_2,  -z_3\right),
\end{equation}
Note that these act in the same way as the generators in \eqref{eq:two-Z2_generators}. 
The orientifold is then generated by the ``standard" anti-holomorphic involution
\begin{equation}
    \sigma (z_1,z_2,z_3)=(\bar{z}_1,\bar{z}_2,\bar{z}_3).
\end{equation}
The cohomology of the resulting space is described by the Hodge numbers $h^{2,1}=h^{1,1}_-=3$ and $h^{1,1}_+=0$. Hence, its geometry can be described in terms of three K\"ahler moduli $T_i$ and four complex-structure moduli $U_i$, which also include the axion-dilaton (see also Eqs \eqref{eq:kdgkt} and \eqref{eq:csdgkt} in Appendix \ref{appendix:DGKT}).\footnote{The K\"ahler and complex structure moduli $\phi_a$ and $\psi_a$ can also be expressed in terms of products and ratios of the radii of the six-torus, respectively.} They can be expanded as
\begin{equation}\label{eq:TU}
    T_i \equiv b_i+ i \e^{\phi_i}\,, \qquad U_a \equiv \xi_a+i \e^{\psi_a},
\end{equation}
where the imaginary parts measure the size of two- and three-cycles in string units, and are analogous to the $\sigma_i$'s (K\"ahler) and $\rho_i$'s (complex structure) in the 3d examples of the previous sections. The real parts instead correspond to $B_2$- and $C_3$-axions respectively, which are not present in 3d due to the orientifold involution. In the language of $\mathcal{N}=1$ supersymmetry, the $\{T_i,U_a \}$ are chiral multiplets, containing both a scalar (saxion) and pseudo-scalar (axion) degree of freedom. The K\"ahler potential can be split as the sum $K = K_\mathrm{K}\left(T_i,\bar{T}_i\right)+K_\mathrm{CS}\left(U_a,\bar{U}_a\right)$ for two sectors, while the inclusion of fluxes generates a polynomial super-potential $W\left(T_i,U_a\right)$. Both are described in detail in Appendix \ref{appendix:DGKT}. At the two-derivative level, the $\mathcal{N}=1$ action is given by
\begin{equation}\label{eq:Ldgkt}
    \mathcal{L}= -2 M_\mathrm{Pl}^2 \sum_{\Phi_I} K^{I \bar{J}} \partial_{\mu} \Phi_I \partial^{\mu}\Phi_{\bar{J}}-2 M_\mathrm{Pl}^4 \e^K \left( \sum_{\Phi_I} K^{I \bar{J}} D_I W D_{\bar{J}} \bar{W}-3|W|^2\right) ,
\end{equation}
where $\Phi_I= \{T_i,U_a \}$ and a non-conventional factor of 2 has been added to make contact with the notation of section \ref{sec:holo_constraint} (see also \cite{Bobev:2025yxp}). From now on, we also omit all factors of $M_\mathrm{Pl}$.

In the $\mathbb{T}^6 / \mathbb{Z}_2^2$ model, the scalar potential \eqref{eq:Ldgkt} admits a supersymmetric vacuum of the DGKT--CFI type described above. All moduli can be stabilised except the three complex structure axions, which remain flat directions at tree-level (they do not appear in either $K$ or $W$). As expected, the spectrum of the low-lying scalars is given by \eqref{eq:deltaK} and \eqref{eq:deltaCS}. In particular, one can go to a basis of canonically normalised scalar fields $\left\{\varphi_i,a_i\right\}_{i=1,...,7}$ for the saxions and the axions respectively, dual to single particle operators of dimensions
\begin{equation}\label{eq:deltaZ2}
    \Delta(\varphi_i) =(10,6,6,6,2,2,2)\,, \qquad \tilde{\Delta} (a_i) =(11,5,5,5,3,3,3)\,.
\end{equation}
Notice that, unlike the three-dimensional examples presented in the previous sections, the axions are not projected out and remain in the spectrum. This has important implications for the calculation of the cubic couplings, as the expansion of the kinetic term \eqref{eq:Ldgkt} typically comes with derivative cubic couplings of the form
\begin{equation}
    \mathcal{L}\supset -\sum_{i,j,k} d_{ijk}\,\varphi_i \partial_{\mu}a_j\partial^{\mu}a_k,
\end{equation}
meaning that the $d_{ijk}$ are not always vanishing in the extremal correlator. With the spectrum in \eqref{eq:deltaZ2}, a variety of extremal and super-extremal arrangements are possible. Because all axions have negative parity, these can only arise for three-point functions of either three saxions or one saxion and two axions. In the next subsections, we analyse these two cases in detail.

\subsubsection*{Super-extremal arrangements for the saxions}
In the pure saxion sector, no derivative couplings arise from the kinetic term (equivalently, $d_{ijk}=0$ identically). Just as in the three-dimensional case, the extremal couplings can be simply read from the couplings in the scalar potential. The spectrum does not allow for extremal arrangements amongst saxions, but does so for super-extremal ones.

The first super-extremal arrangement corresponds to operators of dimension (10,6,2), for which the cubic couplings vanish identically:
\begin{equation}
    c_{1jk}=0\,, \qquad (d_{ijk}=0 )\,, \qquad j=2,3,4\,, \qquad k=5,6,7,
\end{equation}
where the indices denote mass eigenstates, ordered as in eq.~\eqref{eq:deltaZ2}.
For the arrangement (10,2,2), however, the cubic couplings do not vanish, and can be read off as
\begin{equation}\label{eq:1022}
    c_{1 jk}= -\frac{55}{\sqrt{13}} \delta_{j k}\,, \qquad (d_{1 jk}=0)\,, \qquad j,k=5,6,7\,.
\end{equation}
Analogously, for the (6,2,2) super-extremal arrangement, one has
\begin{equation}\label{eq:622}
    c_{i j k}= -\left(\frac{6}{\sqrt{5}} \delta_{i,2} + \frac{2}{\sqrt{5}} \delta_{i,3}+\frac{2}{\sqrt{13}} \delta_{i,4}\right) \delta_{j k}\,, \qquad (d_{i j k}=0)\,, \qquad i=2,3,4 \quad j,k=5,6,7.
\end{equation}
The appearance of such couplings can be traced back to the appearance of certain terms in the potential, since the $d_{ijk}$'s vanish. To allow for a clearer geometric interpretation, paralleling that of section \ref{sec:AdS3IIB_cubic_couplings}, it is convenient to express them in terms of the undiagonalised fields, i.e. those appearing in the definitions \eqref{eq:TU} of the moduli $T_i$ and $U_a$. Then, the extremal couplings \eqref{eq:1022}-\eqref{eq:622} can be shown to arise from
\begin{equation}
   L_\mathrm{AdS}^2 V_\mathrm{cubic} \supset -\frac{55}{208} \left(4\Phi-\Psi\right)\left[4(\psi_1^2+\psi_2^2+\psi_3^2+\psi_4^2)-(\psi_1+\psi_2+\psi_3+\psi_4)^2 \right]
\end{equation}
and
\begin{equation}
   L_\mathrm{AdS}^2 V_\mathrm{cubic} \supset \frac{3}{52} \left(\Phi+3\Psi\right)\left[4(\psi_1^2+\psi_2^2+\psi_3^2+\psi_4^2)-(\psi_1+\psi_2+\psi_3+\psi_4)^2 \right]
\end{equation}
in the original potential, where $\Phi\equiv\phi_1+\phi_2+\phi_3$ and $\Psi \equiv \psi_1+\psi_2+\psi_3+\psi_4$ are not mass eigenstates. Therefore, they can only vanish if one is able to impose the identification
\begin{equation}
    4(\psi_1^2+\psi_2^2+\psi_3^2+\psi_4^2)=(\psi_1+\psi_2+\psi_3+\psi_4)^2.
\end{equation}
From Lagrange's identity, it follows that the only solution is given by 
\begin{equation}\label{eq:psif}
\psi_1=\psi_2=\psi_3=\psi_4 \equiv \psi,    
\end{equation}
i.e. all complex structure moduli must be projected out of the spectrum, and then $\psi$ parametrises the lower-dimensional dilaton. 

\subsubsection*{(Super-)extremal arrangements for one saxion and two axions}
Taking axions into account, many more (super-)extremal arrangements are possible. Some of those are the ones that appear in the original $\mathbb{T}^6/\mathbb{Z}_3^2$ orientifold for DGKT: in particular, they are given by the triples $(10,5,5),\,(11,6,5)$, and the corresponding cubic couplings do also vanish here for the  $\mathbb{T}^6/\mathbb{Z}_2^2$ DGKT--CFI construction, thanks to a non-trivial cancellation. In addition, some new super-extremal arrangements are given by $(3,6,11), \, (3,5,10)$ and $(2,5,11)$. These give rise to vanishing super-extremal couplings, as $c_{ijk}=d_{ijk}=0$ identically. However, there are also some non-vanishing extremal couplings. The first ones come from the extremal arrangement $(6,3,3)$, and are given by
\begin{equation}
    c_{i jk}=0\,, \qquad d_{ijk}= \left(\frac{2}{\sqrt{5}} \delta_{i2} +\frac{2}{3\sqrt{5}}\delta_{i3}+\frac{2}{3\sqrt{13}} \delta_{i4}\right) \delta_{jk}\,, \quad i=2,3,4 \quad j,k=5,6,7\,.
\end{equation}
They are proportional to the couplings in \eqref{eq:622}, likely due to a supersymmetric Ward identity. In terms of the non-diagonalised fields, they originate from the terms
\begin{equation}
    \mathcal{L} \supset -\frac{1}{52}\left(\Phi+3\Psi\right)\left(4 \sum_{j=1}^4\partial_{\mu} \xi_j \partial^{\mu} \xi_j - \partial_{\mu} \Xi \partial^{\mu} \Xi\right),
\end{equation}
where $\Xi= \xi_1+\xi_2+\xi_3+\xi_4$. Similarly, the extremal couplings for the $(2,5,3)$ arrangement do not vanish either, and they are given by
\begin{equation}
    c_{i jk}=0\,, \qquad d_{ijk}=\delta_{ik} \left( \frac{2}{\sqrt{5}}\delta_{j2}+\frac{2}{3\sqrt{5}}\delta_{j3}+\frac{2}{3\sqrt{13}}\delta_{j4} \right)\,, \quad j=2,3,4 \quad i,k=5,6,7.
\end{equation}
These arise from the terms
\begin{equation}
\mathcal{L} \supset
    \frac{1}{40} \sum_{i=1}^4 \psi _i \Big[ \left( \partial_{\mu} \Xi-4 \partial_{\mu}\xi_i\right) \left(4 \partial_{\mu}b_1+\partial_{\mu} \Xi\right)-10 \sum_{j=1}^4\partial_{\mu} \xi_j \partial^{\mu} \xi_j+40 \partial_{\mu} \xi_i \partial^{\mu} \xi_i  \Big].
\end{equation}
Finally, there is a non-zero super-extremal arrangement, corresponding to $(10,3,3).$ Its cubic couplings are 
\begin{equation}
     c_{ijk}=0, \qquad d_{1jk}= \frac{\delta_{j,k}}{\sqrt{13}}\,, \quad j,k=5,6,7\,,
\end{equation}
and they arise from
\begin{equation}
    \mathcal{L} \supset -\frac{1}{208}\left(4\Phi-\Psi\right)\left(4 \sum_{j=1}^4\partial_{\mu} \xi_j \partial^{\mu} \xi_j - \partial_{\mu} \Xi \partial^{\mu} \Xi\right)\,.
\end{equation}
As in the previous section, we see that for all extremal couplings to vanish, the only possibility is given by
\begin{equation}\label{eq:chif}
    \xi_1=\xi_2=\xi_3=\xi_4 \equiv \xi,
\end{equation}
together with eq.~\eqref{eq:psif}. 

\subsection{A larger orbifold group to the rescue}

Starting from the DGKT--CFI construction on the orbifold $\mathbb{T}^6/\mathbb{Z}_2^2$, we have seen how the holographic constraint for extremal operators is satisfied by the K\"ahler moduli, as happens in the $\mathbb{Z}_3^2$ orbifold. However, it is not always satisfied when complex structure moduli are involved. In analogy with the previous section, one might wonder if it is possible to cure the inconsistency by considering a larger orbifold group. Naturally, the $\mathbb{Z}_3^2$ orbifold group does so, but we want to consider also a group which contains the original $\mathbb{Z}_2^2$ as a subgroup and further identifies certain directions in field space. From this point of view, it is quite natural to suggest $\mathbb{Z}_4^2 \supset \mathbb{Z}_2^2$ as a possible candidate.

Indeed, from Eqs. \eqref{eq:psif} and \eqref{eq:chif}, the cubic couplings can only vanish upon imposing the identification (for both real and imaginary parts)
\begin{equation}\label{eq:ide}
   U_1=U_2=U_3=U_4\equiv U,
\end{equation}
where the $\left\{ U_i \right\}$ parametrise the complex structure moduli and the axio-dilaton. This amounts to projecting out all of the complex structure moduli and axions. From a geometric perspective, it admits the interpretation of performing an additional orbifold quotienting, with respect to the larger group $\mathbb{Z}_4 \times \mathbb{Z}_4 \supset \mathbb{Z}_2 \times \mathbb{Z}_2 $.\footnote{The resulting effective theory of the untwisted sector is also identical to the one obtained for the orbifold $\mathbb{T}^6/\mathbb{Z}_3^2$, on which the original construction was based.} The former is generated by the two elements
\begin{equation}\label{eq:Z4xZ4}
     \lambda\left(\, z_1, z_2, \, z_3\right) = \left(i\,z_1, z_2,   -i\, z_3\right), \qquad
     \kappa\left( z_1,  z_2,  z_3\right) = \left(z_1 ,i\, z_2,  -i\,z_3\right)\,,
\end{equation}
whose respective squares give rise to the generators in \eqref{eq:Z2xZ2}. Notice how quotienting with respect to either  $\mathbb{Z}_4 \times \mathbb{Z}_2$ or $\mathbb{Z}_2 \times \mathbb{Z}_4$ is not enough, as it would amount to only a partial identification in \eqref{eq:ide}, i.e. $U_1=U_2$ and $U_3=U_4$.

As in the 3d vacua, it is tempting to look for a connection with a bulk supergravity interpretation in terms of O-plane intersection. For simplicity, let us consider a square fundamental domain, with $z_i \sim z_i+1$ and $z_i\sim z_i+i$. In the $\mathbb{Z}_2^2$ case, the O-planes passing through the singularity at $z_i=0$ are located at
\begin{align}\label{eq:OZ2Z2}
\begin{split}
        &\Im(z_3)=0\,, \quad \Re(z_{1,2})=0\;;  \qquad \Im(z_2)=0\,, \quad \Re(z_{1,3})=0\;;\\
        &\Im(z_1)=0\,, \quad \Re(z_{2,3})=0\;; \qquad  \Im(z_{1,2,3})=0\,.  
\end{split}
\end{align}
The four O-planes intersect pair by pair along a circle, and wrap cycles in different homology classes. In the $\mathbb{Z}_4 \times \mathbb{Z}_4$ case, the new group elements, when combined with the involution, generate many more O-planes through the origin. However, we can prevent this by adding shifts to the generators (satisfying \eqref{eq:invc}), one example goes as follows:
\begin{equation}
    \lambda\left(\, z_1, z_2, \, z_3\right) = \left(i\,z_1, z_2+ \varsigma,   -i\, z_3\right), \qquad
     \kappa\left( z_1,  z_2,  z_3\right) = \left(z_1 + \varsigma,i\, z_2,  -i\,z_3 + \varsigma\right)\,,
\end{equation}
where $\varsigma = (1 + i)/2$. This is one of the choices of shifts that allow only for the same orientifold planes as in the $\mathbb{Z}_2 \times \mathbb{Z}_2$ case.
Again, all the O-planes intersect pairwise along a circle. However, a key difference with respect to the previous case is that they now all wrap cycles in the same homology class, and that now many of the orientifold images belong to the same orbit, reducing the amount of independent O6-planes. However, resolving the orbifold singularities can be complicated, because the $\mathbb{Z}_4^2$ orbifold contains both codimension‑6 and codimension‑4 singularities, and these intersect each other. It is not clear whether a consistent resolution can be achieved. This could be interesting to investigate further in the future. Hence, the $\mathbb{Z}_3^2$ orbifold provides a cleaner way to satisfy the constraint.

\subsection{Remarks on scale-separated vacua of massless type IIA}
Finally, we remark that the same conclusions can be drawn for the scale-separated AdS$_4$ solution in massless type IIA of ref.~\cite{Cribiori:2021djm}. These solutions were inspired by a double T-duality of the DGKT--CFI solutions on $\mathbb{T}^6/\mathbb{Z}_2^2$. Under this specific change of T-dualities, the six-torus becomes a nilmanifold, known as the Iwasawa manifold, and the K\"ahler moduli transform solely into K\"ahler moduli, and complex structure moduli into complex structure moduli. The scalar potential is invariant under T-duality (after some relabelling), so the analysis in this section for massive IIA on $\mathbb{T}^6/\mathbb{Z}_2^2$ is also valid for massless type IIA on $\mathcal{M}_\mathrm{Iwasawa}/\mathbb{Z}_2^2$. We therefore conclude that the solution on the $\mathbb{Z}_{2}^2$ orbifold does not satisfy the constraint, whereas the $\mathbb{Z}_3^2$ and $\mathbb{Z}_4^2$ orbifolds do. We also note that the nilmanifold is characterised by the following Maurer-Cartan equation
\begin{equation}\label{eq:Iwasawa_MC}
    \dd \zeta^{3} = \omega\;  \bar{\zeta}^1\wedge\bar{\zeta}^2\,, \qquad \dd \zeta^{1,2}=0. 
\end{equation}
where the $\zeta^i$'s are complex one-forms, such that $J= i \delta_{ij}\zeta^i  \wedge \bar\zeta^j /2$ and $\Omega = \zeta^1 \wedge \zeta^2 \wedge \zeta^3$. Note that eq.~\eqref{eq:Iwasawa_MC} is invariant under the two $\mathbb{Z}_4$ generators, and the same is true for the standard $\mathbb{Z}_3$ generators, so that indeed we can orbifold the Iwasawa manifold by these orbifold groups as well.
It would be interesting to see what this implies for the M-theory uplift of this solution as well \cite{Cribiori:2021djm,VanHemelryck:2024bas}.

\section{Discussion}\label{sec:discussion}
In this paper, we investigated the holographic consistency of AdS vacua whose putative duals feature integer-valued conformal dimensions of the scalar operators. They easily allow (super)-extremal arrangements, i.e. triplets of scalar operators whose conformal dimensions satisfy $\Delta_i + \Delta_j = \Delta_k - 2n$, $n \in \mathbb{N}$. As argued in ref.~\cite{Bobev:2025yxp}, the cubic couplings \eqref{eq:Cijk} should vanish between extremal scalars, and we tested this explicitly in a known class of AdS$_3$ and AdS$_4$ compactifications arising from $\mathbb{Z}_2^k$ orbifolds. In type IIB, these are the 3d scale-separated solutions on $\mathrm{Nil}_3 \times \mathbb{T}_4$ with two sets of O5-planes, and two classes of 3d solutions on a solvmanifold, one with a single set and another with four sets of O5-planes. In massive IIA, these are 3d solutions on a seven-torus with three sets of O6-planes.
Similarly, we evaluated the cubic couplings for the 4d DGKT--CFI vacua on the $\mathbb{Z}_2^2$ orbifold. 
In all cases, some of the cubic couplings corresponding to the (super-)extremal arrangements do not vanish, and thus violate the holographic constraint of ref.~\cite{Bobev:2025yxp}.
However, we showed that in all cases, except for one of the solvmanifold solutions, the cubic coupling constraint is satisfied when different orbifold groups are considered, effectively projecting out the offending scalars. For the cases in three dimensions, these orbifold groups are non-abelian, and for the 4d DGKT--CFI setup, we studied the $\mathbb{Z}_4^2$ orbifold. All these new orbifold groups have $\mathbb{Z}_2^3$ (for 3d)  and $\mathbb{Z}_2^2$ (for 4d) subgroups, which have the advantage that the cohomology of the new orbifolds is contained in the old ones. Hence, the solutions to the supersymmetry equations and equations of motion are the same as before, upon identifying some of the flux parameters and scalars such that they are consistent with the larger orbifold group.
For the scale-separated solution on $\mathrm{Nil}_3 \times \mathbb{T}_4$, we studied the new non-abelian shift orbifold, isomorphic to $D_4\times \mathbb{Z}_2$, in detail. The orbifold only admits singularities in the four-torus, and there are only two distinct O5-planes.
Surprisingly, we observed that the holographic constraint is typically violated in the $\mathbb{Z}_2^k$ orbifolds due to the O-planes that wrap cycles in distinct homology classes.

\textit{This seems to suggest that the cubic coupling constraint is only satisfied when O-planes wrap cycles in only one homology class.}

\noindent Whenever the orientifold planes wrap cycles in different homology classes, it is not clear whether a resolution of the orbifold singularities would remove O-plane intersections, which is required for a smooth manifold \cite{Junghans:2023yue, Montero:2024qtz}. It is surprising that the holographic constraint seems to know about this, and we intend to investigate this striking connection more deeply in the future. This naturally raises the question whether the scale-separated AdS$_3$ solutions with non-integer conformal dimensions \cite{Farakos:2020phe,Apers:2022zjx, VanHemelryck:2025qok,Miao:2025rgf,Cribiori:2026caf} should be revisited in this light. Although they are not subject to the holographic constraint, the orientifold planes in these constructions still wrap cycles in distinct homology classes, which may indicate a related obstruction. Additionally, the analysis for the 4d DGKT--CFI vacua seems to suggest that the cubic coupling constraint is violated in the presence of geometric complex structure moduli, so whenever the Calabi-Yau is not rigid with $h^{2,1}\neq 0$. We will show this in an upcoming paper. 

Next, in refs.~\cite{Apers:2025pon,Apers:2026lgi}, the authors investigate some of the features that the brane dual of DGKT would have. However, the internal manifold is treated in a coarse-grained manner. Our analysis suggests that the details of the internal manifold are important for obtaining a consistent holographic dual. It would therefore be interesting to see how our observations align with the results of refs.~\cite{Apers:2025pon,Apers:2026lgi} (see also \cite{Bedroya:2025ltj}).

Finally, holographic constraints appear to offer a powerful alternative to more direct methods of diagnosing the consistency of AdS compactifications. Moreover, we have shown in this work that they may be able to admit a microscopic, bulk interpretation in line with suspected obstructions to scale separation. We hope this perspective may be exploited further and ultimately guide us to the construction of field theory dual pairs of scale-separated vacua.

\section*{Acknowledgements}
We would like to thank Ivano Basile, Alexandre Belin, Nikolay Bobev, Mateo Galdeano, Daniel Junghans, Vincent Menet, Hynek Paul, Andrea Sangiovanni, Alessandro Tomasiello, Thomas Van Riet and Farah Verbeure for insightful discussions. FR acknowledges support from a junior postdoctoral fellowship of the Fonds Wetenschappelijk Onderzoek (FWO), project number 12A1Q25N. The work of VVH is supported by 
Kungliga Fysiografiska sällskapet i Lund.

\appendix
\section{Details on the \texorpdfstring{AdS$_3$}{AdS3} compactifications on (co-)closed \texorpdfstring{$G_2$}{G2}-structures in type II string theory}\label{app:AdS3-solutions}
The AdS$_3$ solutions that are discussed in the main text were originally found in \cite{VanHemelryck:2025qok,Farakos:2025bwf} (see also \cite{Arboleya:2024vnp} for the non-supersymmetric cousins of the solvmanifold construction with one set of O5-planes). They are all compactifications on $G_2$-structures, and in the absence of warping and non-trivial dilaton profile, which is the case for smeared sources, supersymmetric solutions to the equations of motion become equivalent to the following conditions in type IIB \cite{Dibitetto:2018ftj,Passias:2019rga,Passias:2020ubv,VanHemelryck:2022ynr}
\begin{equation}\label{eq:G2_torsion}
    \dd \Phi = W_1 \star \Phi + W_{27}\,, \qquad \dd \star \Phi =0\,, \qquad  W_1 = \frac{12\mu}{7}\,,
\end{equation}
where $\mu = L_\mathrm{AdS}^{-1}$ is the inverse AdS radius, and the RR-fluxes are given by
\begin{gather}\label{eq:IIB_RR_sol}
    g_s F_3 = - \star \dd \Phi - 2\mu \Phi\,, \qquad g_s F_7 = 2\mu \vol_7 \\
    F_1 =0\,, \qquad F_5 = 0\,, \qquad H= 0\,.
\end{gather}
Additionally, the non-trivial Bianchi identity of $F_3$ must be solved:
\begin{equation}
    \dd F_3 = j_\mathrm{O5}\,.
\end{equation}
In type IIA, the supersymmetry conditions are equivalent to 
\begin{equation}\label{eq:IIA_G2}
    \dd \Phi = 0\,, \qquad H \wedge \Phi = 0\,,
\end{equation}
and the RR-fluxes must satisfy 
\begin{equation}\label{eq:IIA_RR_sol}
    g_s F_0 = -4\mu\,, \qquad F_2 = 0\,, \qquad g_s F_{4} = -\star H - 2\mu \star \Phi\,, \qquad F_6 = 0\,.
\end{equation}
There are non-trivial Bianchi identities for both $F_2$ and $F_6$:
\begin{equation}
    0 = \dd F_2 = H \wedge F_0 + j_\mathrm{O6}\,, \qquad 
    0 = \dd F_6 = H \wedge F_4 + j_\mathrm{D2} + j_\mathrm{O2}\,.
\end{equation}
When the underlying $G_2$-structure space is an orbifold of a group manifold, such as twisted tori, then the conditions above become algebraic when expanded in a basis like \eqref{eq:basis_Z23_forms}, which then can be easily solved.

It was shown in \cite{VanHemelryck:2022ynr} that these conditions are equivalent to minimising a real superpotential $P$, whose expression appeared before in the literature \cite{Farakos:2020phe,Emelin:2021gzx}, which can be constructed using the $G_2$-structure data and the fluxes. The scalar potential, appearing after dimensional reduction of the 10d action, can also be constructed from the superpotential $P$. For the orbifolds of our twisted tori at hand, these become model-dependent functions in the radii $L_i$ and the lower-dimensional dilaton $\delta_3$, which is defined as 
\begin{equation}
    \delta_3^{-2} = \e^{-2\phi} L_1 L_2 L_3 L_4 L_5 L_6 L_7\,.
\end{equation}
Note that the radii are expressed in string units.
The field space metric, obtained from the kinetic terms, is 
\begin{equation}
    g_{ij} = 2\;\mathrm{diag}\left(4 \delta_3^{-2}\,, L_1^{-2}\,, L_2^{-2}\,, L_3^{-2}\,, L_4^{-2}\,, L_5^{-2}\,, L_6^{-2}\,, L_7^{-2} \right)\,.
\end{equation}
With this, the scalar potential is written in terms of the superpotential $P$:
\begin{equation}
    V = 4 \partial_i P g^{ij} \partial_j P - 4 P^2\,.
\end{equation}
The potential enters the 3d action as follows:
\begin{equation}
    S = \frac{M_\mathrm{Pl}}{2} \int \dd^3 x \sqrt{-g_3} \left(R_3 - \frac{1}{2}g_{ij}\partial_\mu \phi^i \partial^\mu \phi^j - V \right)\,.
\end{equation}
Here, the potential has been defined in such a way that it has mass dimension 2 (instead of 3), and in the vacuum, it satisfies $V_0 = -2 L_\mathrm{AdS}^{-2}$.
In the following subsections, we present the scalar potentials (rather than their superpotentials) for the setups that we have considered in the main text. 
In the rest of this appendix, we normalise the scalars canonically as follows:
\begin{equation}
\label{eq:canonical_scalars}
    L_i = \e^{\sqrt{2}\sigma_i}\,, \qquad \delta_3 = \e^{\Phi_3/\sqrt{2}}\,.
\end{equation}
The mass matrix to diagonalise is then just given by the Hessian of the potential computed with derivatives in $\delta_3$ and the scalars $\sigma_i$. We also explicitly write the scalar field transformations that diagonalise this mass matrix, as those are relevant for the cubic coupling constraint.

\refstepcounter{subsection}
\subsection*{\thesubsection\quad Scale-separated $\mathrm{AdS_3} \times (\mathrm{Nil}_3 \times \mathbb{T}^4)/\mathbb{Z}_2^3$}\label{app:AdS3_Nil3_T4}
This scale-separated setup was first considered in ref.~\cite{VanHemelryck:2025qok}, with the Maurer-Cartan equations for the nilmanifold $\mathfrak{n}_2 = \mathrm{Nil}_3 \times \mathbb{T}^4$ as in \eqref{eq:MC_Nil3_T4}, RR-fluxes as in \eqref{eq:F3-flux}, and appropriate O5-planes to cancel the tadpoles. Solving the supersymmetry conditions \eqref{eq:G2_torsion}-\eqref{eq:IIB_RR_sol}, we find
\begin{align}
    L_{i} &= l_i \left|\frac{\prod_{j<k} F^2_{ijk}}{\prod_{l<m<n}F_{lmn}}\right|^{\frac{1}{4}}\,, \qquad l_i = \{\sqrt{2}\,, \sqrt{2}\,, 1\,, 1\,, 1\,, 1\,, 1/\sqrt{2}\}\\
    g_s &= \frac{\omega}{6}\frac{f_7^{1/2}F_{347}F_{567}}{|\prod_{l<m<n}F_{lmn}|^{1/2}}\,.
\end{align}
The products above involve only the flux parameters appearing in eq.~\eqref{eq:F3-flux}.
Using the techniques in \cite{Emelin:2021gzx,VanHemelryck:2022ynr,VanHemelryck:2025qok}, we can construct the superpotential and consequently the scalar potential, which we write in terms of the radii and 3d dilaton as follows:
\begin{multline} \label{eq:V_Nil3xT4}
    V = \frac{\delta_3^4 }{4 L_1^2 L_2^2 L_3 L_4 L_5 L_6 L_7} \biggl(\delta_3^2 L_1 L_2 (f_7^2 + 
      F_{567}^2 L_1^2 L_2^2 L_3^2 L_4^2 + F_{347}^2 L_1^2 L_2^2 L_5^2 L_6^2 + 
      F_{127}^2 L_3^2 L_4^2 L_5^2 L_6^2\\
      + F_{246}^2 L_1^2 L_3^2 L_5^2 L_7^2 + 
      F_{136}^2 L_2^2 L_4^2 L_5^2 L_7^2 + F_{145}^2 L_2^2 L_3^2 L_6^2 L_7^2 + 
      F_{235}^2 L_1^2 L_4^2 L_6^2 L_7^2)\\
    -   2 \delta_3 L_1 L_2 (F_{567} L_3 L_4 + F_{347} L_5 L_6) L_7 \sqrt{
    L_1 L_2 L_3 L_4 L_5 L_6 L_7} \omega + L_3 L_4 L_5 L_6 L_7^3 \omega^2\biggr)\,.
\end{multline}
We can calculate the masses and the mass eigenstates by diagonalising the Hessian of the potential, expressed in terms of the canonical scalars \eqref{eq:canonical_scalars}. Doing this exercise, we obtain that the scalars $\{\sigma_{3,4,5,6}\}$ are already mass eigenstates, and that the four eigenstates can be expressed in terms of $\{\sigma_{1,2,7},\Phi_3\}$ as follows:
\begin{equation}
    \begin{bmatrix}
        \chi_1 \\ \chi_2 \\ \chi_3 \\  \xi
    \end{bmatrix} = 
    \begin{bmatrix}
        -\frac{1}{\sqrt{2}} & \frac{1}{\sqrt{2}} & 0 &
   0 \\
 \frac{1}{\sqrt{6}} & \frac{1}{\sqrt{6}} &
   \sqrt{\frac{2}{3}} & 0 \\
 -\frac{1}{2 \sqrt{21}} & -\frac{1}{2 \sqrt{21}}  & \frac{1}{2 \sqrt{21}} & \frac{3}{2} \sqrt{\frac{3}{7}}
   \\
 \frac{3}{2 \sqrt{7}} & \frac{3}{2 \sqrt{7}} &
   -\frac{3}{2 \sqrt{7}} & \frac{1}{2 \sqrt{7}} \\
    \end{bmatrix}\cdot
    \begin{bmatrix}
        \sigma_1 \\ \sigma_2 \\ \sigma_7 \\ \Phi_3
    \end{bmatrix}\,.
\end{equation}
These scalars have the following masses and dimensions:
\begin{alignat}{2}
    &m^2(\xi) L_\mathrm{AdS}^2 = 120\,, \quad& &\Delta(\xi) = 12 \\
    &m^2(\chi_{1,2,3}) L_\mathrm{AdS}^2 = 8\,, \quad& &\Delta(\chi_{1,2,3}) = 4 \\
    &m^2(\sigma_{3,4,5,6}) L_\mathrm{AdS}^2 = 8\,, \quad& &\Delta(\sigma_{3,4,5,6}) = 4\,.
\end{alignat}
The cubic couplings we are interested in are obtained from Taylor expanding the potential in the mass eigenstates $\xi$, $\{\chi_{1,2,3}\}$ and $\{\sigma_{3,4,5,6}\}$ at cubic order, for which we need to compute three derivatives of the potential. We will not show the full result, but only present the relevant
terms. Extremal arrangements are not possible with these scaling dimensions, but super-extremal ones of the type (12,4,4) are. We find indeed that they are not vanishing. In AdS units, we find that they are
\begin{equation}
    L_\mathrm{AdS}^2 V_\mathrm{cubic} \supset \frac{1}{3!} 162 \sqrt{\frac{2}{7}} \;\tilde{\xi} \; (\tilde \sigma_3+ \tilde \sigma_4 - \tilde \sigma_5 -\tilde \sigma_6)^2\,,
\end{equation}
as reported in the main text. We note that actually, only the terms involving the flux parameters $F_{347}$ and $F_{567}$ in \eqref{eq:V_Nil3xT4} are responsible for this. These terms encode both the flux contributions to the potential (proportional to $F_{347}^2$ and $F_{567}^2$) and the O5-plane contributions (proportional to $F_{347}\omega$ and $F_{567}\omega$, for which the tadpole condition has been used). 

\refstepcounter{subsection}
\subsection*{\thesubsection\quad AdS$_3$ solutions on solvmanifolds with four sets of O5-planes}\label{app:AdS3_Solv7_4O5}
The AdS$_3$ solutions on a solvmanifold defined by the Maurer-Cartan equations~\eqref{eq:MC_Solv7} and with four sets of O5-planes, were also found in ref.~\cite{VanHemelryck:2025qok}. For this setup, one has $L_1=L_2$, $L_3 = L_4$, $L_5 = L_6$ and solving eqns.~\eqref{eq:G2_torsion}-\eqref{eq:IIB_RR_sol} gives
\begin{equation}
    L_i = \left(\frac{f_7 f_{3,i}^2}{f_{3,1}f_{3,3} f_{3,5}}\right)^{\frac{1}{4}}\,, \qquad L_7 = \frac{\left(f_7 f_{3,1}f_{3,3} f_{3,5}\right)^{\frac{1}{4}}}{2 \tilde{f}_3}\,, \qquad g_s = \frac{2 \omega_\Sigma}{3}\left(\frac{f_7}{f_{3,1}f_{3,3} f_{3,5}}\right)^{\frac{1}{2}}\,,
\end{equation}
for $i=1,3,5$ and $\omega_\Sigma \equiv \omega_1+\omega_3+\omega_5$. The scalar potential can be constructed by the procedure outlined above, and is given by:
\begin{multline}\label{eq:V_Solv7_4O5}
    V =\delta_3^6 \biggl(\frac{f_{7}^2}{4
    L_1  L_2  L_3  L_4  L_5
    L_6  L_7} +\frac{  f_{3,1}^2  L_3  L_4
    L_5  L_6}{4  L_1  L_2
    L_7}+\frac{  f_{3,3}^2  L_1
    L_2  L_5  L_6}{4  L_3  L_4
    L_7}+\frac{  f_{3,5}^2  L_1
    L_2  L_3  L_4}{4  L_5  L_6
    L_7}\\
    +\frac{  \tilde{f}_3^2  L_1
    L_4  L_6  L_7}{4  L_2  L_3
    L_5}+\frac{  \tilde{f}_3^2  L_2
    L_3  L_6  L_7}{4  L_1  L_4
    L_5}+\frac{\tilde{f}_3^2  L_2
    L_4  L_5  L_7}{4  L_1  L_3
    L_6}+\frac{\tilde{f}_3^2  L_1
    L_3  L_5  L_7}{4  L_2  L_4
    L_6}\\
    -\frac{\delta_3^{-1} \omega_\Sigma  \tilde{f}_3}{2}\sqrt{ L_1  L_2  L_3
    L_4  L_5  L_6  L_7} \left[\frac{ 1}{L_1
    L_3  L_5  L_7}+ \frac{1}{L_2  L_4  L_5
    L_7}+\frac{1}{L_2  L_3
    L_6  L_7}+\frac{1}{L_1
    L_4  L_6  L_7}\right]\\
    +\frac{\delta_3^{-2}}{4 L_7^2} \left[
   \omega_1^2 \left(\frac{L_2^2}{L_1^2} + \frac{L_1^2}{ L_2^2} \right)+\omega_3^2 \left(\frac{L_4^2}{L_3^2} + \frac{L_3^2}{ L_4^2} \right)+\omega_5^2 \left(\frac{L_6^2}{L_5^2} + \frac{L_5^2}{ L_6^2} \right)-2(\omega_1^2+\omega_3^2+\omega_5^2) \right] \biggr)\,.
\end{multline}
Expressing the potential in terms of the canonical fields and diagonalising its Hessian, leads to the following mass eigenstates
\begin{equation}
\begin{bmatrix}
    \zeta\\
    \chi_1\\
    \chi_2\\
    \chi_3\\
    \chi_4\\
    \rho_1\\
    \rho_3\\
    \rho_5
\end{bmatrix}=
\frac{1}{\sqrt{10}}\begin{bmatrix}
 0 & 0 & 0 & 0 & 0 & 0 & 3 &
   1 \\
 0 & 0 & 0 & 0 & 0 & 0 & -1 &
   3 \\
 0 & 0 & 0 & 0 & \sqrt{5} & \sqrt{5}
   & 0 & 0 \\
 0 & 0 & \sqrt{5} & \sqrt{5} & 0 & 0
   & 0 & 0 \\
 \sqrt{5} & \sqrt{5} & 0 & 0 & 0 & 0
   & 0 & 0 \\
 0 & 0 & -\sqrt{5} & \sqrt{5} & 0 &
   0 & 0 & 0 \\
 -\sqrt{5} & \sqrt{5} & 0 & 0 & 0 &
   0 & 0 & 0 \\
 0 & 0 & 0 & 0 & -\sqrt{5} &
   \sqrt{5} & 0 & 0 
\end{bmatrix} \cdot
\begin{bmatrix}
    \sigma_1\\
    \sigma_2\\
    \sigma_3\\
    \sigma_4\\
    \sigma_5\\
    \sigma_6\\
    \sigma_7\\
    \Phi_3\\
\end{bmatrix}\,,
\end{equation}
which have the following masses and scaling dimensions for the dual operators:
\begin{gather}
    m^2(\zeta)L_\mathrm{AdS}= 3\,, \qquad \Delta(\zeta) = 3\\
    m^2(\chi_{i})L_\mathrm{AdS}= 8\,, \qquad \Delta(\chi_{i}) = 4\\
    m^2(\rho_{j})L_\mathrm{AdS}= -1 + 36 \frac{\omega_j^2}{\omega_\Sigma^2}\,, \qquad \Delta(\rho_{j}) = 1+6 \left| \frac{\omega_j}{\omega_\Sigma}\right|\,.
\end{gather}
Extremal and super-extremal arrangements can only occur when at least one of the $\{\rho_j\}$ scalars is involved. To see whether they appear, we Taylor expand the potential at cubic order, and look at terms that only involve $\{\rho_j\}$. Denoting $\Delta_j \equiv \Delta(\rho_j)$, we find
\begin{multline}\label{eq:cubicV_Solv7_4O5}
     (3!)L_\mathrm{AdS}^2 V_\mathrm{cubic} \supset -42 \tilde\rho_1 \tilde\rho_3 \tilde\rho_5 + \frac{24}{\sqrt{5}} \; \tilde\chi_1 \left[ \tilde\rho_1^2 (-2 +\Delta_1) \Delta_1 + \tilde\rho_3^2 (-2 + \Delta_3) \Delta_3 + \tilde\rho_5^2 (-2 +  \Delta_5) \Delta_5\right]\\
     -\frac{3}{\sqrt{5}}\tilde\zeta \left[\tilde\rho_1^2(-5 + 2 \Delta_1)(1+2 \Delta_1)+\tilde\rho_3^2(-5 + 2 \Delta_3)(1+2 \Delta_3)+\tilde\rho_5^2(-5 + 2 \Delta_5)(1+2 \Delta_5)\right]\,.
\end{multline}
We now explore for which values of the conformal dimensions the cubic coupling constraint invalidates the solution. We do this for each term:
\begin{enumerate}
    \item $\tilde \rho_1 \tilde \rho_3 \tilde\rho_5$: It is clear that any triple $(\Delta_1, \Delta_3, \Delta_5)$ leading to (super-)extremal arrangements is not allowed.
    \item $\tilde\chi_1 \tilde\rho_i^2$: We see that they are extremal $\Delta_i + \Delta_i = 4$ if $\Delta_i =2$, for which the cubic couplings vanish precisely. For the super-extremal arrangement $\Delta_i + \Delta_i = 4 - 2 \times 1$ with $\Delta_i = 1$, the couplings do not vanish. This cannot occur when all the $\omega_i \neq 0$.
    \item $\tilde \zeta \tilde \rho_i^2$: They are extremal $\Delta_i + \Delta_i = 3$ if $\Delta_i = 3/2$. The cubic couplings do not vanish in this case. So $\Delta_i = 3/2$ is forbidden.
\end{enumerate}
As remarked in the main text, we also see that if all the scalars $\rho_1$, $\rho_3$ and $\rho_5$ are projected out, the cubic couplings vanish automatically. We remark that 
\begin{equation}
    \e^{2\rho_1} = \frac{L_2}{L_1}\,, \qquad \e^{2\rho_3} = \frac{L_4}{L_3}\,, \qquad \e^{2\rho_5} = \frac{L_6}{L_5}\,,
\end{equation}
so an orbifold fixing $L_1=L_2$, $L_3=L_4$ and $L_5 = L_6$ realises this. We discuss such an orbifold in appendix \ref{app:orbifold_Solv7}.
Finally, we remark also that the cubic couplings in \eqref{eq:cubicV_Solv7_4O5} arise only from the terms depending on the flux parameter $\tilde{f}_3$ in \eqref{eq:V_Solv7_4O5}, including flux potential and O5-plane contributions (proportional to $\tilde{f}_3^2$ and $\tilde{f}_3 \omega$ respectively).

\refstepcounter{subsection}
\subsection*{\thesubsection\quad AdS$_3$ solutions on solvmanifolds with one set of O5-planes}\label{app:AdS3_Solv7_1O5}
The analysis for the solvmanifold compactification, where only one set of O5-planes is needed, is very similar. The Maurer-Cartan equations \eqref{eq:MC_Solv7} are the same, but the RR-fluxes are now given by \eqref{eq:RR_Solv7_1O5}. Here, we must choose $-\omega_1=\omega_3=\omega_5$. The solutions to supersymmetry equations \eqref{eq:G2_torsion}-\eqref{eq:IIB_RR_sol} are such that $L_{i+1} = \sqrt{2} L_i$ for $i = 1,3,5$, and that the remaining scalars are given by
\begin{equation}
    L_i = \left(\frac{f_7 f_{3,i}^2}{2f_{3,1}f_{3,2} f_{3,3}}\right)^{\frac{1}{4}}\,, \qquad L_7 = \frac{\left(f_7 f_{3,1}f_{3,3} f_{3,5}\right)^{\frac{1}{4}}}{2^{\frac{1}{4}} \tilde{f}_3}\,, \qquad g_s = \omega_1\left(\frac{f_7}{2f_{3,1}f_{3,2} f_{3,3}}\right)^{\frac{1}{2}}\,.
\end{equation}
These are also the critical points of the scalar potential, which can again be constructed by the procedure outlined above, which is given by
\begin{multline}
   V = \delta_3^6 \biggl(\frac{ f_7^2}{4
    L_1  L_2  L_3  L_4  L_5
    L_6  L_7}+ \frac{f_{3,1}^2  L_3  L_4
    L_5  L_6}{4  L_1  L_2
    L_7}+\frac{ f_{3,3}^2  L_1
    L_2  L_5  L_6}{4  L_3  L_4
    L_7}+\frac{ f_{3,5}^2  L_1
    L_2  L_3  L_4}{4  L_5  L_6
    L_7}\\
    +\tilde{f}_{3}^2\left[\frac{    L_1
    L_4  L_6  L_7}{4  L_2  L_3
    L_5}+\frac{    L_2
    L_3  L_6  L_7}{4  L_1  L_4
    L_5}+\frac{    L_2
    L_4  L_5  L_7}{4  L_1  L_3
    L_6}+\frac{  L_1
    L_3  L_5  L_7}{ L_2  L_4
    L_6}\right]\\
    -\frac{3  \delta_3^{-1} \tilde{f}_{3}
   \omega_1 \sqrt{ L_1  L_2  L_3
    L_4  L_5  L_6  L_7}}{2  L_1
    L_3  L_5  L_7}\\
    +\delta_3^{-2}\omega_1^2\left[\frac{ 
    L_2^2}{4  L_1^2 L_7^2}+\frac{L_1^2}{4  L_2^2L_7^2}+\frac{L_4^2}{4  L_3^2 L_7^2}+\frac{L_3^2}{4  L_4^2 L_7^2}+\frac{L_6^2}{4 L_5^2 L_7^2}+\frac{L_5^2}{4  L_6^2 L_7^2}-\frac{3}{2 L_7^2}\right]\biggr)\,.
\end{multline}
Expressing this potential in canonical variables and diagonalising its Hessian leads to the following mass eigenstates:
\begin{equation}
\begin{bmatrix}
    \xi_1\\
    \xi_2\\
    \xi_3\\
    \chi_1\\
    \chi_2\\
    \chi_3\\
    \chi_4\\
    \theta\\
\end{bmatrix}=
\begin{bmatrix}
 \frac{1}{2} & -\frac{1}{2} & 0 & 0 & -\frac{1}{2}
   & \frac{1}{2} & 0 & 0 \\
 \frac{1}{2 \sqrt{3}} & -\frac{1}{2 \sqrt{3}} &
   -\frac{1}{\sqrt{3}} & \frac{1}{\sqrt{3}} &
   \frac{1}{2 \sqrt{3}} & -\frac{1}{2 \sqrt{3}} & 0
   & 0 \\
 \frac{\sqrt{\frac{5}{3}}}{4} &
   -\frac{\sqrt{\frac{5}{3}}}{4} &
   \frac{\sqrt{\frac{5}{3}}}{4} &
   -\frac{\sqrt{\frac{5}{3}}}{4} &
   \frac{\sqrt{\frac{5}{3}}}{4} &
   -\frac{\sqrt{\frac{5}{3}}}{4} & \frac{3
   \sqrt{\frac{3}{5}}}{4} &
   \frac{\sqrt{\frac{3}{5}}}{4} \\
 0 & 0 & 0 & 0 & 0 & 0 & -\frac{1}{\sqrt{10}} &
   \frac{3}{\sqrt{10}} \\
 0 & 0 & 0 & 0 & \frac{1}{\sqrt{2}} &
   \frac{1}{\sqrt{2}} & 0 & 0 \\
 0 & 0 & \frac{1}{\sqrt{2}} & \frac{1}{\sqrt{2}} &
   0 & 0 & 0 & 0 \\
 \frac{1}{\sqrt{2}} & \frac{1}{\sqrt{2}} & 0 & 0 &
   0 & 0 & 0 & 0 \\
 -\frac{1}{4} & \frac{1}{4} & -\frac{1}{4} &
   \frac{1}{4} & -\frac{1}{4} & \frac{1}{4} &
   \frac{3}{4} & \frac{1}{4} \\
\end{bmatrix} \cdot
\begin{bmatrix}
    \sigma_1\\
    \sigma_2\\
    \sigma_3\\
    \sigma_4\\
    \sigma_5\\
    \sigma_6\\
    \sigma_7\\
    \Phi_3\\
\end{bmatrix}\,.
\end{equation}
The scalars have the following masses and scaling dimensions:
\begin{gather}
    m^2(\xi_i)L_\mathrm{AdS}^2 = 48\,, \qquad \Delta(\xi_i) = 8\\
    m^2(\chi_j)L_\mathrm{AdS}^2 = 8\,, \qquad \Delta(\chi_j) = 4\\
    m^2(\theta)L_\mathrm{AdS}^2 = 0\,, \qquad \Delta(\theta) = 2\,,
\end{gather}
with $i=1,2,3$ and $j=1,2,3,4$.
The potential dangerous extremal arrangements are of the type (4,2,2) and (8,4,4), with corresponding cubic couplings of the schematic form $\theta \theta \chi$ and $\chi \chi \xi$. All such couplings are vanishing. This is similar for the super-extremal arrangements of the type (8,4,2) corresponding to cubic couplings of the form $\xi \chi \theta$.
However, for the super-extremal arrangements of the type (8,2,2), corresponding to cubic couplings of the form $\theta^2 \xi$, this is no longer the case. Indeed, we see that
\begin{equation}
    L_\mathrm{AdS}^2 V_\mathrm{cubic} \supset  \frac{1}{3!} 24 \sqrt{30}\;\tilde{\xi}_3 \;\tilde{\theta}^2\,.
\end{equation}
Furthermore, we note that
\begin{equation}
    \e^{4\frac{\sqrt{10}}{3} \;\xi_3} = \delta_3^2 L_7^3 \left(\frac{L_1 L_3 L_7}{L_2 L_4 L_6}\right)^{5/3}\,, \qquad \e^{\sqrt{8}\; \theta} =  \delta_3^2 L_7^3 \frac{L_2 L_4 L_6}{L_1 L_3 L_5} = \left(g_s \frac{L_7}{L_1 L_3 L_5}\right)^2\,.
\end{equation}
The non-trivial dilaton dependence in both scalars does not allow us to consider another orbifold to project them out, and the cubic coupling constraint invalidates this solution.

\refstepcounter{subsection}
\subsection*{\thesubsection\quad AdS$_3$ solutions in massive type IIA on tori}\label{app:AdS3_T7}
In the case of massive type IIA, the solution of interest with integer conformal dimensions was found in \cite{Farakos:2025bwf}. This is a solution on an orbifold of a $\mathbb{T}^7$, with fluxes as in eq.~\eqref{eq:T7_fluxes} and necessary sources.
They are solutions to the supersymmetry equations \eqref{eq:IIA_G2}-\eqref{eq:IIA_RR_sol}, and these solutions are given by 
\begin{gather}
    L_1^2 = \frac{f_{43}f_{44}}{h_3 h_5} g_s^2\,, \quad L_2^2 = \frac{f_{41}f_{42}}{h_3 h_5} g_s^2\,, \qquad L_3^2 = \frac{f_{41}f_{44}}{h_1 h_5} g_s^2\,, \quad L_4^2 = \frac{f_{42}f_{43}}{h_1 h_5} g_s^2\,,\\ \label{eq:IIA_T7_sol}
    L_5^2 = \frac{f_{42}f_{44}}{h_1 h_3} g_s^2\,, \quad L_6^2 = \frac{f_{41}f_{43}}{h_1 h_3} g_s^2\,, \qquad L_7^2 =\frac{4 (h_1 h_3 h_5)^2}{F_0^2 f_{41}f_{42}f_{43} f_{44}}g_s^{-6}\,.
\end{gather}
We note that full moduli stabilisation is not possible in this case, which is reflected by the fact that the radii above are still expressed in terms of the string coupling.
These solutions can also be found by minimising the following scalar potential:
\begin{multline}
    V=\delta_3^6 \biggl(\frac{1}{4}   f_0^2  L_1  L_2
    L_3  L_4  L_5  L_6
    L_7+\frac{  f_{44}^2  L_2
    L_4  L_6}{4  L_1  L_3  L_5
    L_7}+\frac{   f_{42}^2  L_1
    L_3  L_6}{4  L_2  L_4  L_5
    L_7}+\frac{ f_{41}^2  L_1
    L_4  L_5}{4  L_2  L_3  L_6
    L_7}+\frac{ f_{43}^2  L_2
    L_3  L_5}{4  L_1  L_4  L_6
    L_7}\\
    -\frac{ \delta_3^{-1}  f_0  h_1
   \sqrt{ L_3 L_4 L_5 L_6}}{2 \sqrt{ L_1 L_2 L_7}}-\frac{ \delta_3^{-1}  f_0
    h_3 \sqrt{ L_1 L_2 L_5 L_6}}{2 \sqrt{ L_3 L_4 L_7}}-\frac{ \delta_3^{-1}
    f_0  h_5 \sqrt{ L_1 L_2 L_3 L_4}}{2 \sqrt{ L_5 L_6 L_7}}\\
   +\frac{ \delta_3^{-2}
    h_1^2}{4  L_1^2  L_2^2
    L_7^2}+\frac{ \delta_3^{-2}  h_3^2}{4
    L_3^2  L_4^2
    L_7^2}+\frac{ \delta_3^{-2}  h_5^2}{4
    L_5^2  L_6^2  L_7^2}\biggr)\,.
\end{multline}
Computing the mass eigenstates, obtained as usual by expressing the potential in terms of the canonical scalars \eqref{eq:canonical_scalars} and diagonalising the Hessian of the potential, one finds
\begin{equation}
\begin{bmatrix}
    \xi\\
    \chi_1\\
    \chi_2\\
    \chi_3\\
    \chi_4\\
    \theta_1\\
    \theta_2\\
    \theta_3\\
\end{bmatrix}=
\begin{bmatrix}
\frac{\sqrt{\frac{3}{5}}}{4} &
   \frac{\sqrt{\frac{5}{3}}}{4} &
   \frac{\sqrt{\frac{5}{3}}}{4} &
   \frac{\sqrt{\frac{5}{3}}}{4} &
   \frac{\sqrt{\frac{5}{3}}}{4} &
   \frac{\sqrt{\frac{5}{3}}}{4} &
   \frac{\sqrt{\frac{5}{3}}}{4} & \frac{3
   \sqrt{\frac{3}{5}}}{4} \\
 -\frac{3}{\sqrt{10}} & 0 & 0 & 0 & 0 & 0 & 0 &
   \frac{1}{\sqrt{10}} \\
 0 & 0 & 0 & 0 & 0 & -\frac{1}{\sqrt{2}} &
   \frac{1}{\sqrt{2}} & 0 \\
 0 & 0 & 0 & -\frac{1}{\sqrt{2}} & \frac{1}{\sqrt{2}} &
   0 & 0 & 0 \\
 0 & -\frac{1}{\sqrt{2}} & \frac{1}{\sqrt{2}} & 0 & 0 &
   0 & 0 & 0 \\
 0 & -\frac{1}{\sqrt{3}} & -\frac{1}{\sqrt{3}} &
   \frac{1}{2 \sqrt{3}} & \frac{1}{2 \sqrt{3}} &
   \frac{1}{2 \sqrt{3}} & \frac{1}{2 \sqrt{3}} & 0 \\
 0 & 0 & 0 & -\frac{1}{2} & -\frac{1}{2} & \frac{1}{2} &
   \frac{1}{2} & 0 \\
 -\frac{1}{4} & \frac{1}{4} & \frac{1}{4} & \frac{1}{4}
   & \frac{1}{4} & \frac{1}{4} & \frac{1}{4} &
   -\frac{3}{4} \\
\end{bmatrix} \cdot
\begin{bmatrix}
    \Phi_3\\
    \sigma_1\\
    \sigma_2\\
    \sigma_3\\
    \sigma_4\\
    \sigma_5\\
    \sigma_6\\
    \sigma_7\\
\end{bmatrix}\,.
\end{equation}
These scalars have the following masses and conformal dimensions:
\begin{gather}
    m^2(\xi)L_\mathrm{AdS}^2 = 48\,, \qquad \Delta(\xi) = 8\\
    m^2(\chi_i)L_\mathrm{AdS}^2 = 8\,, \qquad \Delta(\chi_i) = 4\\
    m^2(\theta_j)L_\mathrm{AdS}^2 = 0\,, \qquad \Delta(\theta_j) = 2\,,
\end{gather}
for $i=1,2,3,4$ and $j = 1,2,3$.
It is important to note that the potential does not depend on the scalar $\theta_3$ at all, and hence has a flat direction, as noticed below eq.~\eqref{eq:IIA_T7_sol}.
Expanding the potential around the vacuum results in no cubic couplings for the extremal arrangements. For the super-extremal arrangements of the type $(8,2,2)$ corresponding schematically to terms $\xi \theta \theta$ in the potential, we find the following:
\begin{equation}\label{eq:cubicV_T7}
   L_\mathrm{AdS}^2 V_\mathrm{cubic} \supset - \frac{1}{3!} 16 \sqrt{30} \; \tilde \xi \; (\tilde \theta_1^2 + \tilde \theta_2^2)
\end{equation}
The scalars $\theta_1$ and $\theta_2$ can be expressed back in terms of the radii:
\begin{equation}
    \e^{\sqrt{24}\;\theta_1} = \frac{L_3 L_4 L_5 L_6}{L_1^2 L_2^2}\,, \qquad \e^{\sqrt{8}\;\theta_2} = \frac{L_5 L_6}{L_3 L_4}\,.
\end{equation}
Again, a new orbifold that sets $L_1 L_2 = L_3 L_4 = L_5 L_6$ would project these scalars out and make such couplings disappear.
Also here, we note that the non-vanishing cubic couplings in \eqref{eq:cubicV_T7} come entirely from the terms in the potential parametrised by the NSNS-fluxes $\{h_{1,3,5}\}$, including the NSNS-flux potentials and the O6-plane contributions (proportional to $\{h_{1,3,5}^2\}$ and $\{h_{1,3,5} \omega\}$ respectively).

\section{Non-abelian orbifolds for \texorpdfstring{AdS$_3$}{AdS3} solutions}\label{app:other_orbifolds}
In this appendix, we first discuss some aspects of shift orbifolds on nilmanifolds, and then we discuss some alternative orbifolds of the solvmanifold and six-torus that cure the cubic coupling constraint by projecting out the relevant moduli.

\refstepcounter{subsection}
\subsection*{\thesubsection\quad Comments on shift orbifolds of nilmanifolds}\label{app:shifts_nil}
As introduced in the main text, we consider a three-dimensional nilmanifold that we have parametrised by the directions 1, 2 and 7. We solve the Maurer-Cartan equations \eqref{eq:MC_Nil3_T4} in terms of coordinates as follows:
\begin{equation}
    e^1 = \dd x^1, \qquad e^2 = \dd x^2, \qquad e^7 = \dd x^7 +\frac{1}{2}x^1 \dd x^2 - \frac{1}{2}x^2 \dd x^1\,.
\end{equation}
Other choices are possible as well, but can be obtained from this one by diffeomorphisms. 
The nilmanifold is made compact by quotienting with a lattice, which here leads to the following lattice identifications:
\begin{equation}
    (x^1, x^2, x^7) \sim \left(x^1+n^1, x^2 +n^2, x^7 + n^7 -\frac{1}{2}n^1 x^2 + \frac{1}{2} n^2 x^1\right)\,.
\end{equation}
As was explained in \cite{Frey:2002hf,Cribiori:2021djm}, the integers $n^i$ should be elements of $2\mathbb{Z}$.
It was also argued earlier that $\mathbb{Z}_2^k$-orbifolds cannot include shifts in the directions that do not correspond to the centre of the algebra, here $x^1$ and $x^2$ \cite{Andriolo:2018yrz}. 
For instance, the map 
\begin{equation}
    (x^1, x^2, x^7) \to (- x^1-1, -x^2, x^7)\,,
\end{equation}
does not leave the one-form $e^7$ invariant, as $e^7 \to e^7 + e^2/2$.
The reason is that this shift does not correspond to half a lattice vector, which is required by eq.~\eqref{eq:orbifold_conditions}.
An easy fix is to respect that condition by adding the corresponding shift in $x^7$. Indeed, for a simple involution, we can take half a lattice vector to shift by, for example:
\begin{equation}
    g_\alpha(x^1, x^2, x^7) =  \left(- x^1-1, -x^2, x^7- \frac{1}{2}x^2\right)\,.
\end{equation}
Note that this is precisely how the orbifold group generator $\alpha$ acts in section~\ref{sec:AdS3IIB_cubic_couplings}.
Similarly, we can consider another involution on these coordinates as follows:
\begin{equation}
    g_\gamma(x^1, x^2, x^7) = \left(- (x^1+1), x^2+1, -\left(x^7+\frac{1}{2}x^1- \frac{1}{2}x^2\right)\right)\,,
\end{equation}
which is how the orbifold element $\gamma$ acts in section~\ref{sec:AdS3IIB_cubic_couplings}. 
We see indeed that these maps square to the identity:
\begin{align}
\begin{split}
    &g_\alpha^2 (x^1, x^2, x^7) = (x^1, x^2, x^7)\,,\\
    &g_\gamma^2(x^1, x^2, x^7) = \left(x^1, x^2+2, x^7+\frac{1}{2}2 x^1\right) \sim (x^1, x^2, x^7)\,,
\end{split}
\end{align}
where the latter is indeed a shift by a lattice vector.
\refstepcounter{subsection}
\subsection*{\thesubsection\quad Orbifold for solvmanifold in type IIB}\label{app:orbifold_Solv7}
In this section, we consider the orbifold that cures the holographic constraint in the solvmanifold compactification of section~\ref{sec:other_AdS3_cubic_couplings}.
The solvmanifolds Maurer-Cartan equations can be satisfied when the co-frame one-forms take the following forms:
\begin{gather}
    e^{i} = \cos (\omega_i x^7) \dd x^i -\sin(\omega_i x^7) \dd x^{i+1}\,,\\
    e^{i+1} = \sin (\omega_i x^7) \dd x^i +\cos(\omega_i x^7) \dd x^{i+1}\,,\quad e^{7} = \dd x^7\,,
\end{gather}
where $i=1,3,5$ and the lattice identifies all coordinates $x^j \sim x^j+1$, and it was argued in \cite{VanHemelryck:2025qok} to look at $\omega_i \in 4\pi \mathbb{Z}$.
Consider the group $\left<\kappa, \lambda, \gamma\right>$, where the generators act on the coordinates as follows:
\begin{align}
    \kappa(x^1,x^2,x^3,x^4,x^5,x^6, x^7) &= \left(x^1+\frac{1}{2},-x^2+\frac{1}{2},x^4,x^3,x^6,x^5, -x^7+\frac{1}{2}\right)\\
    \lambda(x^1,x^2,x^3,x^4,x^5,x^6, x^7) &= \left(x^2,x^1,x^4+\frac{1}{2},x^3+\frac{1}{2},x^5,-x^6, -x^7+\frac{1}{2}\right)\\
    \gamma(x^1,x^2,x^3,x^4,x^5,x^6, x^7) &= \left(-x^1,x^2,-x^3,x^4,-x^5+\frac{1}{2},x^6+\frac{1}{2}, -x^7+\frac{1}{2}\right)
\end{align}
These generators leave the one-forms and Maurer-Cartan equations \eqref{eq:MC_Solv7} invariant.
Notice that none of the generators commutes and that $\kappa^2 = \lambda^2 =\gamma^2= 1$. Moreover, they satisfy
\begin{align}
&\kappa \lambda \kappa \lambda = \lambda \kappa \lambda \kappa = \alpha\\
&\kappa \gamma \kappa \gamma = \gamma \kappa \gamma \kappa = \beta\\
&\lambda \gamma \lambda \gamma = \gamma \lambda \gamma \lambda = \alpha \beta = \beta\alpha\,.
\end{align}
with $\alpha$, $\beta$ and $\gamma$ as in eq.~\eqref{eq:three-Z2_generators}.
This means that this orbifold has a $\mathbb{Z}_2^3$ subgroup generated by $\left<\alpha,\beta,\gamma\right>$, and that again, the cohomology of the new orbifold is contained in the cohomology of the original $\mathbb{Z}_2^3$ orbifold.
The generators also satisfy the following relations:
\begin{equation}
    \Theta^a \Theta^b \Theta^c= \Theta^c \Theta^b \Theta^a \quad  \Leftrightarrow \quad \Theta^a \Theta^b \Theta^c  \Theta^a = \Theta^c \Theta^b\,, \qquad \Theta^{a} \in \{\kappa,\lambda,\gamma\}\,,
\end{equation}
which is valid for any combination of the group elements. The group has 32 elements, and all of them require only up to five generators to be generated.
For this orbifold, we note that only the following combinations of three-forms are preserved
\begin{equation}
    e^{127}\,, \quad e^{347}\,, \quad e^{567}\,, \quad e^{145}+e^{136}+ e^{235}-e^{246}\,.
\end{equation}
This enforces $L_1=L_2$, $L_3=L_4$ and $L_5=L_6$, and the left-invariant three-form becomes:
\begin{equation}
    \Phi = L_1^2 L_7 e^{127}+ L_3^2 L_7 e^{347} + L_5^2 L_7 e^{567} + L_1 L_3 L_5 \left( e^{145}+e^{136}+e^{235}-e^{246}\right)\,.
\end{equation}
We then consider an orientifold involution $\sigma_\gamma$ that acts in the same manner as $\gamma$ without shifts. Note that all generators send $x^7 \to -x^7 +1/2$.
By design, $\sigma_\gamma g$ has no fixed points if $g$ consists of an odd number of generators, as it would send $x^7 \to x^7+1/2$. If $g$ consisted of an even number of generators, $\sigma_\gamma g$ just sends $x^7 \to - x^7$ and thus the only O5-planes present should be localised in the coordinate $x^7$, which is what we want. The additional shifts ensure that no O3-involutions are created. We do not discuss the full singularity structure of the orbifold and orientifold and leave this for future research.

We remark that we can also consider other orbifolds that project out the required scalars. One other example is a non-abelian orbifold $\left< T, Q, \gamma\right>$, where $T$ and $Q$ generate a $\mathbb{Z}_3^2$ group, as in DGKT \cite{DeWolfe:2005uu}. This group is non-abelian, as the generator $\gamma$ as in $\eqref{eq:three-Z2_generators}$ does not commute with either $T$ or $Q$, and a version of this (with shifts) appeared in ref.~\cite{MR1787733}, sec.~12.8.4. Although it has no $\mathbb{Z}_2^3$ subgroup, it would lead to the same cohomology as in the orbifold above. A necessary caveat is that this orbifold requires a different lattice for the compact space, which we do not investigate further.

\refstepcounter{subsection}
\subsection*{\thesubsection\quad Orbifold for seven-torus in type IIA}\label{app:orbifold_T7}
In this section, we present a non-abelian orbifold group that cures the cubic coupling constraint for the compactification in massive IIA of ref.~\cite{Farakos:2025bwf}, as described in sec.~\ref{sec:mIIA_AdS3_T7}.
We consider the group $\left<\alpha,\eta,\gamma\right>$, where the generators act as follows:
\begin{align}
    &\alpha( x^1, x^2,x^3,x^4,x^5,x^6,x^7) = (-x^1,-x^2, x^3, x^4, -x^5, -x^6,x^7)\\
    &\eta( x^1, x^2,x^3,x^4,x^5,x^6,x^7) = (x^3,x^4,x^5,x^6,x^1,x^2, x^7)\,,\\
    &\gamma(x^1, x^2,x^3,x^4,x^5,x^6,x^7) = \left(-x^1,x^2, -x^3, x^4, -x^5, x^6,-x^7+\frac{1}{2}\right)\,.
\end{align}
Note that $\eta^3 = 1$ and that $\gamma$ commutes with both $\alpha$ and $\eta$, but $\alpha$ and $\eta$ do not commute, such that $\eta \alpha \eta\eta = \beta$, with $\beta$ acting as in eq.~\eqref{eq:three-Z2_generators}. So also this orbifold has the original $\mathbb{Z}_2^3$ group as a subgroup. The group has 24 elements in total, and it has the property that 
\begin{gather}
    \alpha \eta \eta  \alpha = \eta  \alpha \eta\,,\\
    \eta  \eta \alpha = \alpha \eta  \alpha \eta\,, \quad \alpha \eta  \eta = \eta  \alpha \eta \alpha\,.
\end{gather}
where the last line can be obtained from the first by multiplying it from the left or right by $\alpha$ or $\eta$. This orbifold group is isomorphic to $A_4 \times \mathbb{Z}_2$ and is also described (with other shifts) in ref.~\cite{MR1787733}, sec.~12.8.3.
Only the following three-forms are invariant
\begin{equation}
    e^{127}+ e^{347}+e^{567}\,, \qquad e^{145}+e^{136}+e^{235}\,, \qquad e^{246}\,,
\end{equation}
which enforces $L_1 = L_3 = L_5$ and $L_2 = L_4 = L_6$, and the $G_2$ left-invariant three-form becomes
\begin{equation}
    \Phi = L_1 L_2 L_7 (e^{127}+ e^{347}+e^{567}) + L_1^2 L_2 (e^{145}+e^{136}+e^{235}) - L_2^3 e^{246}\,.
\end{equation}
Finally, it turns out that an O2-involution
\begin{equation}
    \sigma( x^1, x^2,x^3,x^4,x^5,x^6,x^7) =( -x^1, -x^2,-x^3,-x^4,-x^5,-x^6,-x^7)\,,
\end{equation}
generates all necessary O6-planes. By design, $\sigma g$ has no fixed points if $g$ contains an odd number of $\gamma$'s, and the fixed points of $\sigma \alpha$, $\sigma \beta$ and $\sigma \alpha \beta$ lead to O6-planes, whereas all other combinations of $\sigma$ with $\alpha$ and $\eta$ lead to further O2 images.  Notice that, if one starts from an O6-involution, an O2-involution comes naturally from the orbifold. For this case, we do not consider the singularity structure of the orbifold and leave that to further research.

\section{Details on the DGKT--CFI \texorpdfstring{AdS$_4$}{AdS4} vacua in type IIA string theory}\label{appendix:DGKT}
In this appendix, we discuss more details of the DGKT--CFI construction on the $\mathbb{T}^6/\mathbb{Z}_2\times \mathbb{Z}_2$ orbifold, with Hodge numbers $h^{2,1}=h^{1,1}_- =3$ and $h^{1,1}_+=0$. From the point of view of the low-energy EFT, this translates into the existence of three Kahler moduli
\begin{equation}\label{eq:kdgkt}
    T_i= b_i + i \e^{\phi_i}, \qquad i=1,2,3,
\end{equation}
where the $e^{\varphi_i}$ are the $2-$cycle volumes and the $b_i$ are $B_2$ axions. Unlike the original $\mathbb{T}^6 / \mathbb{Z}_3^2$ example, in this case, there are also three complex structure moduli. They are denoted by
\begin{equation}\label{eq:csdgkt}
    U_a= \xi_a+i \e^{\psi_i}, \qquad i=1,2,3,
\end{equation}
where the $\xi_a$ are $C_3$ axions. The K\"ahler potential is given by $K=K_\mathrm{K}+K_Q$, where the contribution of the K\"ahler sector is
\begin{equation}
    K_\mathrm{K} = -\log \left(8 \kappa \frac{T_1-\bar{T}_1}{2i} \frac{T_2-\bar{T}_2}{2i} \frac{T_3-\bar{T}_3}{2i} \right)\,, 
\end{equation}
and the complex structure contribution is
\begin{equation}
    K_Q=-\log \left( P(\e^{\psi_1},\e^{\psi_2},\e^{\psi_3},\e^{\psi_4}) \right)\,.
\end{equation}
The latter is expressed in terms of the function $P(x_i)$, which plays the role of a pre-potential. It is a homogeneous polynomial of degree $h^{2,1}+1$, and in this particular case it is given by
\begin{equation}
    P(x_1,x_2,x_3,x_4)= 2^4 x_1 x_2 x_3 x_4.
\end{equation}
The complex structure moduli are also related to the 4d dilaton as
\begin{equation}
P(\e^{\psi_1},\e^{\psi_2},\e^{\psi_3},\e^{\psi_4}) = \e^{-4 D},
\end{equation}
so that one also has $K_Q=4D$. In this basis, the kinetic terms then read
\begin{equation}
    \mathcal{L}_\mathrm{kin}=-\frac{1}{2} \sum_{i=1}^3 \left( \partial_{\mu} \phi_i \partial^{\mu} \phi_i+\e^{-2 \phi_i} \partial_{\mu} b_i \partial^{\mu} b_i\right) -\frac{1}{2} \sum_{a=1}^4 \left( \partial_{\mu} \psi_a \partial^{\mu} \psi_a+\e^{-2 \psi_a} \partial_{\mu} \xi_a \partial^{\mu} \xi_a \right).
\end{equation}
The inclusion of non-zero fluxes generates a superpotential, which allows to stabilise all moduli except the complex structure moduli. In particular, 
\begin{equation}
F_0=m_0, \quad F_2=m^a \omega_a, \quad F_4=e_a \widetilde{\omega}^a, \quad F_6= e_0, \quad H_3 = p^k \beta_k
\end{equation}
where the $\omega_a,\widetilde{\omega}^a$ are a basis of $H_-^{1,1}$ and $H_-^{2,2}$ respectively, and the $\beta_k$ a basis of $H_-^{2,1}$. In absence of D6 branes, the tadpole condition implies
\begin{equation}
    m_0 \sum_{k=1}^{h^{2,1}+1} p_k<0\,.
\end{equation}
For simplicity, we assume $m_0>0$ and $p_k <0$, $k=1,...,h^{2,1}+1$. Then, the superpotential takes the form
\begin{equation}
     W=\sum_{i=1}^3 e_i T_i-2 \sum_{i=1}^4 U_a p_a+e_0+\kappa \left(m_3 T_1 T_2+m_1 T_3 t_2+m_2 T_1 T_3\right)- \kappa m_0 T_1 T_2 T_3.
\end{equation}
Without loss of generality, we can set $m_a=0$. The scalar potential is given by
\begin{equation}\label{eq:Vdgkt}
\begin{split}
    V=& \frac{\e^{-\Phi-\Psi} \sum _{i=1}^3 \e^{2 \phi _i} \left(\frac{\kappa  m_0 b_1 b_2 b_3 }{b_i}-e_i\right){}^2}{32 \kappa }-\frac{3}{32} \kappa  m_0^2 \e^{-\Psi+\Phi} \\ & +\frac{\kappa ^2 m_0^2 \e^{-\Psi+\Phi} \sum _{i=1}^3 b_i^2 \e^{-2 \phi _i}}{32 \kappa } +\frac{\e^{-\Phi-\Psi} \sum _{i=1}^4 \left(p_i \e^{\psi _i}+\frac{1}{2} \kappa  m_0 \e^{\Phi}\right){}^2}{8 \kappa }\\ & +\frac{\e^{-\Psi-\Phi} \left(b_1 e_1+b_2 e_2+b_3 e_3-b_1 b_2 b_3 \kappa  m_0+e_0-2 \xi _1 p_1-2 \xi _2 p_2-2 \xi _3 p_3-2 \xi _4 p_4\right){}^2}{32 \kappa},
    \end{split}
\end{equation}
where we have defined $\Phi=\phi _1+\phi _2+\phi _3$ and $\Psi=\psi _1+\psi _2+\psi _3+\psi _4$. If the signs of the fluxes are chosen to ensure $\rm{sgn} (m_0 e_i)=-1$, the potential \eqref{eq:Vdgkt} admits a supersymmetric minimum when the saxions take the values
\begin{equation}\label{eq:mins}
    \phi_i =\log \left( \frac{\bar{v}}{|e_i|}\right), \quad i=1,2,3; \qquad  \psi_a = \log \left( \frac{\bar{v}}{|p_a|}\right), \quad a=1,2,3,4,
\end{equation}
with $\bar v$ satisfying
\begin{equation}
    \bar{v}= \sqrt{\frac{5}{3} \left| \frac{e_1 e_2 e_3}{k m_0}\right| }.
\end{equation}
The axion vev is given by
\begin{equation}\label{eq:mina}
    b_i =0\,, \quad i=1,2,3\;; \quad \quad p_1\xi_1+p_2\xi_2+p_3\xi_3+p_4\xi_4= \frac{e_0}{8}\,,
\end{equation}
with three exactly flat directions. The vacuum energy takes the value 
\begin{equation}
    V |_{\rm{min}}= -\frac{243}{200} \sqrt{\frac{3}{5}} m_0 p_1 p_2 p_3 p_4 \left|\frac{k m_0}{e_1 e_2 e_3}\right|^{3/2}.
\end{equation}
To derive the spectrum and the interactions, one can expand \eqref{eq:Vdgkt} in the fluctuations around the minimum \eqref{eq:mins}-\eqref{eq:mina}, $\delta \phi_i, \delta \psi_a$ for the saxions and $\delta b_i, \delta \xi_a$ for the axions. To arrive at canonically normalised mass matrices, one can perform the field redefinition
\begin{equation}
    \begin{pmatrix}
        \delta \phi_i\\
        \delta \psi_a\\
    \end{pmatrix} = U \, \Bigg( 
   \varphi_j \Bigg)
    ;\qquad
    \begin{pmatrix}
        \delta b_i\\
        \delta \xi_a\\
    \end{pmatrix} = U \, \Bigg(a_j\Bigg)
, \qquad j=1,...,7\,,
\end{equation}
where $U$ is the orthogonal, $7 \times 7$ matrix
\begin{equation}
    U = \left(
\begin{array}{ccccccc}
 -\frac{2}{\sqrt{13}} & \frac{1}{\sqrt{5}} & -\frac{4}{3 \sqrt{5}} & -\frac{4}{3 \sqrt{13}} & 0 & 0 & 0 \\
 -\frac{2}{\sqrt{13}} & 0 & 0 & \frac{3}{\sqrt{13}} & 0 & 0 & 0 \\
 -\frac{2}{\sqrt{13}} & 0 & \frac{\sqrt{5}}{3} & -\frac{4}{3 \sqrt{13}} & 0 & 0 & 0 \\
 \frac{1}{2 \sqrt{13}} & \frac{1}{\sqrt{5}} & \frac{1}{3 \sqrt{5}} & \frac{1}{3 \sqrt{13}} & -\frac{1}{\sqrt{2}} & -\frac{1}{\sqrt{6}} & \frac{1}{2 \sqrt{3}} \\
 \frac{1}{2 \sqrt{13}} & \frac{1}{\sqrt{5}} & \frac{1}{3 \sqrt{5}} & \frac{1}{3 \sqrt{13}} & 0 & 0 & -\frac{\sqrt{3}}{2} \\
 \frac{1}{2 \sqrt{13}} & \frac{1}{\sqrt{5}} & \frac{1}{3 \sqrt{5}} & \frac{1}{3 \sqrt{13}} & 0 & \sqrt{\frac{2}{3}} & \frac{1}{2 \sqrt{3}} \\
 \frac{1}{2 \sqrt{13}} & \frac{1}{\sqrt{5}} & \frac{1}{3 \sqrt{5}} & \frac{1}{3 \sqrt{13}} & \frac{1}{\sqrt{2}} & -\frac{1}{\sqrt{6}} & \frac{1}{2 \sqrt{3}} \\
\end{array}
\right).
\end{equation}
This translates to the conformal dimensions
\begin{equation}
    \Delta_{\varphi} =(10,6,6,6,2,2,2) \qquad \Delta_{a} =(11,5,5,5,3,3,3)\,,
\end{equation}
for the saxions and axion respectively.

\bibliographystyle{JHEP}
\bibliography{refs.bib}

\end{document}